\documentclass[final,3p,times]{elsarticle}

\usepackage{amssymb}
\usepackage{color, colortbl}
\definecolor{light-gray}{gray}{0.95}
\definecolor{medium-gray}{gray}{0.87}
\definecolor{dark-gray}{gray}{0.79}

\definecolor{gray1}{gray}{0.95}
\definecolor{gray2}{gray}{0.9}
\definecolor{gray3}{gray}{0.85}
\definecolor{gray4}{gray}{0.8}
\definecolor{gray5}{gray}{0.75}
\definecolor{gray6}{gray}{0.7}
\usepackage{multirow}
\usepackage{rotating} 

\usepackage{array}
\newcolumntype{?}{!{\vrule width 1pt}}

\newlength{\Oldarrayrulewidth}
\newcommand{\Cline}[2]{
  \noalign{\global\setlength{\Oldarrayrulewidth}{\arrayrulewidth}}
  \noalign{\global\setlength{\arrayrulewidth}{#1}}\cline{#2}
  \noalign{\global\setlength{\arrayrulewidth}{\Oldarrayrulewidth}}
}

\usepackage{tabu}

\begin{document}

\begin{frontmatter}



\title{Human-Collective Visualization Transparency}


\author[label1]{Karina A. Roundtree}
\author[label2]{Jason R. Cody}
\author[label3]{Jennifer Leaf}
\author[label1]{H. Onan Demirel}
\author[label3]{Julie A. Adams}

\address[label1]{School of Mechanical, Industrial, and Manufacturing Engineering, Oregon State University}
\address[label2]{Department of Electrical Engineering and Computer Science, United States Military Academy}
\address[label3]{Collaborative Robotics and Intelligent Systems Institute, Oregon State University}

\begin{abstract}
Interest in collective robotic systems has increased rapidly due to the potential benefits that can be offered to operators, such as increased safety and support, who perform challenging tasks in high-risk environments. Human-collective transparency research has focused on how the design of the algorithms, visualizations, and control mechanisms influence human-collective behavior. Traditional collective visualizations have shown all of the individual entities composing a collective, which may become problematic as collectives scale in size and heterogeneity, and tasks become more demanding. Human operators can become overloaded with information, which will negatively affect their understanding of the collective’s current state and overall behaviors, which can cause poor teaming performance. An analysis of visualization transparency and the derived visualization design guidance, based on remote supervision of collectives, are the primary contributions of this manuscript. The individual agent and abstract visualizations were analyzed for sequential best-of-n decision-making tasks involving four collectives, composed of 200 entities each. The abstract visualization provided better transparency by enabling operators with different individual differences and capabilities to perform relatively the same and promoted higher human-collective performance. 
\end{abstract}




\end{frontmatter}


\section{Introduction}

Complex human-collective robotic interface design must leverage fundamental human factors visualization principles in order to aid accurate perception and comprehension of information that informs operator actions. Transparency, the principle of providing easily exchangeable information to enhance comprehension \cite{Roundtree2019}, can help attain meaningful and insightful information exchanges between an operator and the collective. Integrating transparency into the human-collective interface design, such as implementing different visualization techniques, can help limit poor operator behaviors and improve the human-collective system's overall effectiveness. Achieving transparency; however, becomes more challenging as the collective robotic system's scale changes, and the amount of information needed increases in order to understand the collective's current, or planned actions. Providing too much, or too little transparency may overload, or underload, respectively, the operator and negatively affect the desired outcomes. Designers of collective robotic systems, which are composed of large groups (i.e., $\ge50$) of simplified individual robotic entities, will experience challenges due to the quantity and quality of the information provided by the individual entities. There is a need to investigate how system components, such as the visualization, impact human-collective behaviors (i.e., human-collective interactions and performance) in order to achieve transparency for realistic collective use scenarios. 

Collective robotic systems exhibit biologically inspired behaviors seen in spatial swarms \cite{Brambilla2013}, colonies \cite{Gordon1999}, or a combination of both \cite{Seeley2010}. A bee colony searching for a new hive location exemplifies collective behavior. Initially a subset of the colony population leaves the hive in search of a new hive for a daughter colony \cite{Seeley2010}. The subset of bees fly to a nearby tree branch, where they wait while scout bees search the surrounding area for a new hive location. During the initial flight, the bees exhibit spatial swarm behaviors, similar to those found in schools of fish \cite{Couzin2005} and flocks of birds \cite{BallerinietalNAS2008}, where each bee maintains a particular distance from their neighbor in order to avoid collisions and follow their neighbor in a particular direction. Scout bees find possible hive locations and evaluate each option with respect to ideal hive criteria. The scouts return to the waiting colony in order to begin a selection process (i.e., colony behavior) entailing debate and building consensus on the best hive location (i.e., best-of-\textit{n} \cite{Valentini2017}). After completing a consensus decision-making process, the bees travel to the new hive location, transitioning from colony based behaviors back to spatial swarm behaviors. Upon arrival at the new hive, the bees transition to colony based behaviors. 

Collective robotic system attributes, such as global intelligence and emergent behaviors, are advantageous for task completion, because collectives are: (1) scalable (i.e., can change in size) \citep{Brambilla2013}, (2) resilient to failures (i.e., responsibilities can be redistributed to other collective entities) \citep{Bayindir2007}, and (3) flexible in varying environments \citep{Hein2015,Komareji2013}, as well as the type of robotic entities (i.e., heterogeneous members). Collectives have many proposed applications to aid human operators in conducting challenging tasks, such as environmental monitoring, exploration, search and rescue missions, infrastructure support, and protection for military applications \cite{Sahin2005}. Adding a human operator, who may possess information that a collective does not, can positively influence the collective's consensus decision-making process by minimizing the time to make decisions and ensuring the collective selects a higher valued option. 

The complexity of human-collective systems challenge designers to determine an appropriate quantity and quality of information necessary to support an accurate perception of the collective's state, comprehension of what the collective is doing currently and why, as well as what, the collective plans to do in the future. Investigations analyzing the system design elements, such as algorithms, visualizations, operator control mechanisms, and the interactions that transpire at the junction of these components are necessary in order to understand how to achieve transparency for human-collective systems. This manuscript analyzed visualization transparency provided to a human supervisor \citep{Scholtz2003} using a traditional collective representation, Individual Agents (IA) \citep{Roundtree20191}, and an abstract Collective \citep{Cody2018} representation. The single human operator-collective system incorporated four hub-based collectives, each tasked with making a series of sequential best-of-\textit{n} decisions \citep{Valentini2017}. Each collective was to choose the best option from a set of \textit{n} options, as described in the honeybee nest site search example \citep{Seeley2010}. The objective was to determine which visualization achieved better transparency by evaluating how the visualized information impacted operators with different individual capabilities, their comprehension, the interface usability, and the human-collective team's performance. Focusing on the visualization is necessary, considering the means of communication and interaction with remote collectives will only occur via an interface. Realistic use scenarios will rely heavily on the operator's visual sensory system. Understanding how the visualization design, such as the use of information windows, icons, or colors to represent different collective states, impacts the operator's ability to positively influence the collective's decision-making process is necessary to inform transparency focused design guidelines for human-collective systems.  

\section{Related Work}

Behaviors of spatial swarms and colonies, which constitute collective behaviors, are provided in order to develop an understanding of what collective characteristics may be important to a human teammate and collective system designers. Understanding how collective entities communicate and interact with one another to influence individual entity and global collective state changes is necessary to ground collective system design. Factors that affect transparency, or are influenced by transparency, such as explainability, usability, and performance, can be measured to assess the visualizations are discussed. Understanding the transparency factors and how they influence the human-collective system is necessary in order to inform design decisions. A review of transparency research focused on designing and evaluating collective visualizations with respect to methods of achieving transparency is presented. 

\subsection{Collective Behavior}

Spatial swarm robotics are inspired by self-organized social animals \cite{Brambilla2013}, such as flocks of birds \cite{Ballerini2008} or schools of fish \cite{Couzin2005}, and exhibit intelligent, emergent behaviors as a unit, by responding to locally available information \cite{Sahin2005}. Spatial swarms rely on distributed, localized, and often implicit communication, in order to spread information across the swarm. Salient movements warning individual entities of a nearby predator \cite{Treherne1981} or rapid changes in acceleration, such as a streaking bee guiding a swarm in a particular direction \cite{Seeley2010}, are examples of spatial swarm communication. Spatial swarms follow basic rules of repulsion, attraction, and orientation that enable individual entities to position themselves relative to neighboring entities \cite{Couzin2002}. Couzin et al.'s model states that individual entities in the zone of repulsion attempt to maintain a minimum distance from their neighbors, striving to avoid collisions, while entities that are far from their neighbors will move closer, as a factor of the zone of attraction. The zone of orientation causes individual entities to align themselves in the same direction as their close proximity neighbors.  

Robotic colonies are decentralized systems \cite{Bonabeau1999} composed of entities that exhibit unique roles, such as foraging, which adapt over time to maintain consistent states in changing conditions \cite{Wilson1984}. Biological colonies use a centralized space to share information, such as inside of a nest, or embed information into the environment, such as ants depositing pheromones to indicate a route from a food site to the nest \cite{Holldobler1990}. Positive feedback loops support gaining a consensus to change the colony's behavior \cite{Sumpter2006}. Recruitment for potential changes increases until a quorum is reached. The colony will transition into a decision-making state, implementing different strategies, such as the bee waggle dance \cite{Seeley2010}, in order to reach a consensus. Once a consensus has been reached colony members will indicate to one another that the colony is ready to move to the new nest site, such as physically picking up ants and transporting them to the new nest site \cite{Mallon2001}. Negative feedback mitigates saturation issues, such as food source exhaustion \cite{Bonabeau1999}.

\subsection{Transparency Factors}

\begin{figure}[bp!]
\centering
	\includegraphics[width=\textwidth]{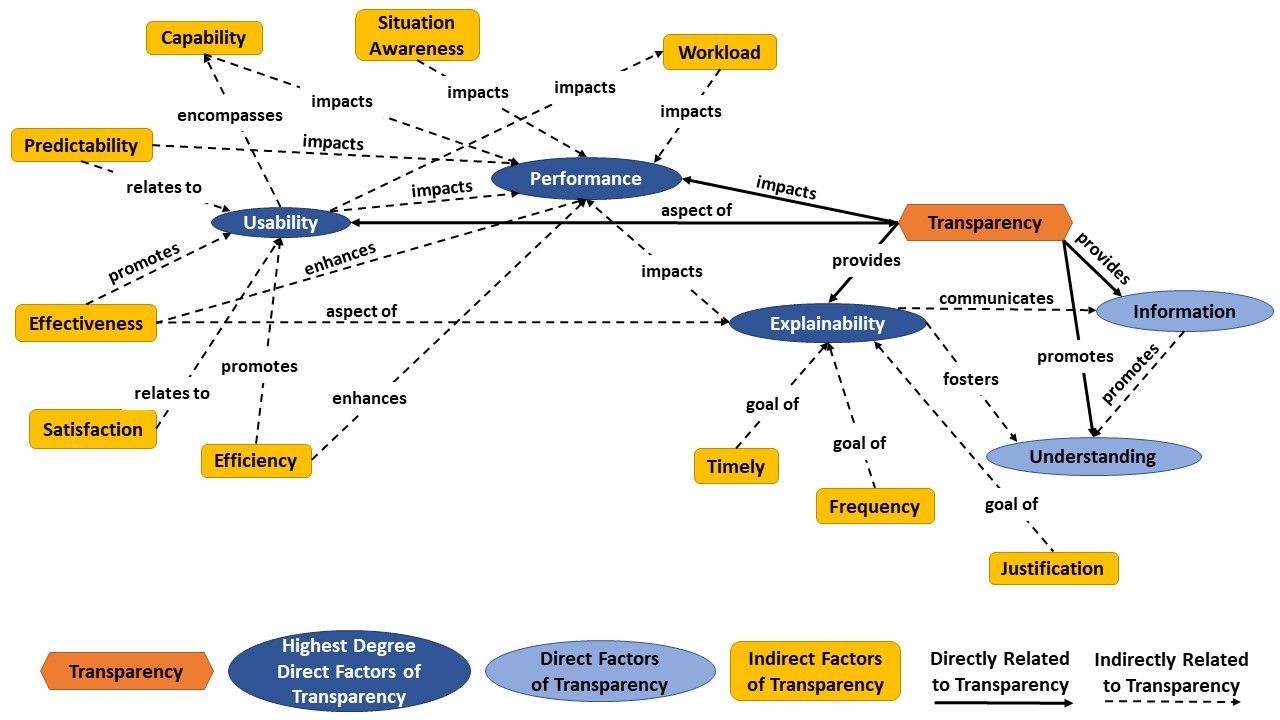}
	\caption{Concept map showing a subset of direct and indirect transparency factors \citep{Roundtree2019} assessed in this human-collective evaluation.}
    \label{fig: Concept Map}
\end{figure}

Transparency has been defined as the communication \cite{Mercado2016}, process \cite{Nedbal2013}, method \cite{LyonsetalAHF2017}, mechanism \cite{Theodorou2017}, property \cite{Selkowitz2015}, emergent characteristic \cite{Ososky2014}, explanation \cite{Kim2006}, and quality \cite{Chenetal2014} of an interface to support operator comprehension of a system's intent. Chen et al.'s transparency definition \cite{Chenetal2014} is commonly used in the robotics domain and uses the three level Situation awareness-based Agent Transparency (SAT) framework, which leverages the human operator trust calibration 3Ps model (purpose, process, and performance) with performance history \cite{Lee2004}, Endsley's situation awareness model \cite{Endsley1995}, and the Beliefs, Desires, and Intentions Agent framework \cite{Rao1995}, in order to achieve transparency.

Transparency for this manuscript builds on Chen et al.'s definition and is defined as the principle of providing information that is easy to use \citep{Kaltenbach2017} in an exchange between human operators' and collectives to promote comprehension \citep{Mark2005,Lyons2013,Alarcon2017}, intent, roles, interactions \citep{ChenetalTIE2018}, performance \citep{Chenetal2014}, future plans, and reasoning processes \citep{Theodorou2017,Helldin2014}. The term principle describes the process of identifying what factors affect and are influenced by transparency, why those factors are important, how the factors may influence one another, and how to design a system to achieve transparency. A subset of transparency factors that originated from human-machine literature \citep{Roundtree2019}, shown in Figure \ref{fig: Concept Map}, will be used to assess the visualization transparency. Five factors impact transparency directly and are represented as blue ovals connected to the transparency node by solid black lines. The three highest total degree (number of in degree + number of out degree) direct factors are explainability, usability, and performance (dark blue). Information and understanding (light blue) are not considered high degree factors, because explainability uses information to communicate and promote understanding. \textit{Explainability} and \textit{usability}, which is a multifaceted quality that influences the operator's perception of a system, are used for the implementation of transparency in the visualization. \textit{Performance} can be used to assess the visualization transparency by determining how well a human-collective team was able to produce an output when executing a task under particular conditions \citep{Alarcon2017}.

The yellow rectangles represent factors that impact visualization transparency indirectly. The timing, frequency, and quantity of information visualized on an interface, such as collective status or feedback, requires consideration of the human operator's capability limitations \citep{Entin1996}, system limitations, as well as task and environmental impacts, in order to be explainable \citep{Atoyan2006}. Human-collective efficiency and effectiveness can be improved by visualizing information, such as predicted collective states, in a clear and cohesive manner that helps alleviate the time and effort an operator must exert to integrate information in order to draw conclusions \citep{StJohn2005} and justify particular actions \citep{Fox2017}. Poor judgements may occur if the operator lacks an understanding of what the collective is doing due to poor visualization usability, which may cause operator dissatisfaction. Visualizations that do not provide needed information to the operator and system control mechanisms to effectively influence collective behavior will hinder human-collective performance, lower operator situational awareness, impact workload negatively, and potentially compromise safety of the human-collective team. Understanding the relationships between the direct and indirect factors, and their relationship to transparency, is needed in order to assess metrics that can quantify good design of human-collective visualizations.

\subsection{Collective Visualization Transparency}

Three design methods that have been used frequently to optimize transparency and desired outcomes (i.e., effective team interactions and high performance) in traditional human-machine systems are: (1) provide system features, such as status, feedback, planning mechanisms, and engagement prompts, (2) use specific guidelines, such as Gestalt principles, to design a system, or (3) train the operators and system \citep{Roundtree2019}. The majority of the collective visualization transparency literature has focused primarily on the first two design methods of providing system features and using specific guidelines to design the collective visualizations. Most evaluations used training, the third method, only to teach operators how to interact with their specific systems \cite{Nagavalli2015,Brown2016}, but did not analyze how different training strategies may impact transparency. More training may be required for different collective visualizations due to difficulties understanding the presented information. Determining what strategies have been used and their associated outcome success rates in remote teaming with collectives, as well as methods that can be evaluated more in depth, such as training, will aid in the development of transparency guidelines for collective visualizations. 

The collective transparency literature that uses traditional collective visualizations (i.e., showing the position of all individual entities) has focused on assessing operator understanding of collective behavior and human-collective performance. A variety of individual collective entity features have been visualized for operators, including current and predicted future position \citep{Walker2012} and heading direction \citep{Brown2016}, health (i.e., speed, strength, capability, and dispersion \citep{Haas2009}), and status. The most commonly used visual icon for an individual collective entity has been a circle, where directional information was observed either as the entity moves across a 2-D space, or the circle also encompassed a line pointing in the heading direction. Some visualization design guidance came from subject matter experts \citep{Haas2009} and the Gestalt principles \citep{Nagavalli2015}. The use of multimodal cues (i.e., visual collective state color coding and written messages, spoken messages, and vibrations) aided operators in the identification and response of signals during a reconnaissance mission (99.9\% accuracy), as well as shorter response times, and lower workload. Sharing collective information via multimodal cues may alleviate an operator's high visual loads when managing multiple tasks on various displays by increasing situational awareness \citep{Haas2009} and promoting better transparency. Operators' using a visualization incorporating Gestalt-based design principles perceived and approximated optimal swarm performance faster than operators using visualizations containing only individual entity position information \citep{Nagavalli2015}. Increased visualization transparency enabled operators to learn when to approximate optimal input timing. 

Information latency, which occurs due to communication bandwidth limitations, and neglect benevolence, the time allowed for a collective to stabilize before issuing new operator commands, on operator understanding of future collective behavior are important considerations \citep{Walker2012}. Latency affected operators' ability to control a collective, but providing additional transparency via a predictive visualization, which showed the predicted location of each individual entity 20 seconds into the future, mitigated these effects. Operators using the predictive visualization with latency performed as well as operators who experienced no latency. Transparency of human-collective systems can be improved by implementing predictive visualizations of the collective and its entities by allowing the operator more time to think about their future actions. Operators will be able to balance span, the number of individual collective entities they interact with, and persistence, the duration of the interactions with visualizations that provide heading information \citep{Brown2016}. Aspects, such as the presence or absence of Couzin et al.’s communication model states, visualized individual collective entities’ velocity, and individual operator characteristics, such as gender, have impacted the identification of swarming behavior \citep{Harvey2018}. Understanding the influence of factors is necessary to promote perception and comprehension of collective behavior to inform future operator actions. 
   
Abstract collective visualizations have been proposed to improve operator understanding and influence positive collective behavior. Radial visualizations, using the SAT model \citep{Chenetal2014} and heuristic evaluations analyzing the application of collective metrics on visualizations \citep{Manning2015}, as well as glyphs \citep{Walker2020}, bounding ellipses \citep{Walker2016}, convex hulls, and directed arrows \citep{Roundtree2018} have been assessed. Similar system features provided in the traditional visualizations were conveyed using the abstract visualizations. The radial visualizations promoted perception (SAT Level 1) by displaying mission \citep{Crandall2017} and collective state information \citep{Ashcraft2019}, such as the direction the entities left the hub to explore environmental targets. Predictions of future collective headings (SAT Level 3), that were provided by elongating the radial display in the direction of more collective support, aided operator actions. Visualizations providing predictive information are needed to achieve transparency for human-collective systems, regardless if the visualization is traditional or abstract. Operators using a glyph were able to acquire information regarding the collective’s power levels, task type, and the number of individual entities, via one icon \citep{Walker2020}. Additional information about particular system features was accessible via pop-up windows. Designers of abstract collective visualizations can ensure transparency by providing redundant information via the collective icon and using supplementary information windows. Conflicting results were found for evaluations assessing whether traditional or abstract visualizations aided operators better during different tasks. Abstract visualizations during a go-to and avoid task in the presence or absence of obstacles \citep{Roundtree2018} performed worse than the traditional visualizations, while abstract visualizations performed as well or better when perceiving biological collective structure \citep{Seiffert2015} and under variable bandwidth conditions \citep{Walker2016}. Further analysis is needed in order to determine which collective visualization will promote better transparency for a common task by investigating how transparency factors influence human-collective behaviors. 
   
\section{Human-Collective Task}

The human-collective task involved a single human operator who supervised and assisted four robotic collectives that performed a sequential best-of-$n$ decision-making task, where the human-collective team chose the best option from a finite set of $n$ options \cite{Valentini2017}. The human-collective team performed two sequential decisions per collective (moved the collective to a new hub site). The decision-making task entailed the identification and selection of the highest valued target within a 500 $m$ range of the current hub, the collective hub moved to the selected target, and initiated the second target selection decision. The consensus decision-making task required a quorum detection mechanism to estimate when the highest valued target was identified by 30\% of the collective \cite{Cody2018}. Each collective of 200 simulated Unmanned Aerial Vehicles searched an urban area of approximately 2 $km^2$. 

The four collective hubs were visible at the start of each trial. Targets became visible as each was discovered by a collective's entities. The target's value was assessed by the collectives' entities, who returned to their respective hub to report the target location and value. The collectives were only allowed to discover and occupy targets within their search range, but some targets were within proximity of multiple hubs. A collective's designated search area changed after it moved to a new target to establish the new hub site. The operator was instructed to prevent multiple collectives merging by not permitting their respective hubs to move to the same target. When a collective moved to a target, the hub moved to the target location, and the target was no longer visible to the operator or available to other collectives. The collective that moved its hub to a target's location first, when two collectives were investigating the same target, moved to the target location, while the second collective returned to its previous location. Both collectives made a decision when a merge occurred, even though only one collective moved its hub to the respective target location.

The general interface design requirements, related to collective autonomy, control, and transparency, are: 1) enable the operator to estimate the collectives' decision-making process, 2) identify appropriate controls to influence the decision-making process, and 3) implement the desired controls \cite{Cody2018}. Two algorithmic models were used. A sequential best-of-\textit{n} decision-making algorithm ($M_{2}$) adapted an existing model, which based decisions on the quality (i.e., value) of a target \cite{Reina2015}. Information exchanges between a collective's entities was restricted inside the hub, to mimic honeybees. Episodic queuing cleared messages when the collective entities transitioned to different states, which resulted in more successful and faster decision completion. Interaction delay and interaction frequency were added as bias reduction methods in order to consider a target's distance from the collective hub and increase interactions among the collectives' entities regarding possible hub site locations. Interaction delay improved the success of choosing the ground truth best targets, and interaction frequency improved decision time. The baseline model ($M_{3}$) allows the collective entities to search and investigate potential targets, but was unable to build consensus. The operator was required to influence the consensus-building element and select that final target, based on the consensus. Simulations were ran without an operator for the $M_{2}$ model in order to understand the operator's influence on collective behavior, referred to as $M_{2 SIM}$. The $M_{3}$ model required operator influence in order to perform the decision-making task; thus, simulation only analysis was not conducted.

The interface controls allowed the operator to alter the collectives' internal states, including their levels of autonomy, throughout the sequential best-of-\textit{n} selection process. The collective's entities were in one of four states. \textit{Uncommitted} entities explored the environment searching for targets, and were recruited by other entities while inside of the collective's hub. Collective entities that \textit{favored} a target reassessed the target's value periodically, and attempted to recruit other entities within the collective's hub to investigate the specified target. Collective entities were \textit{committed} to a particular target once a quorum of support was detected, or after interacting with another committed entity. \textit{Executing} collective entities moved from the collective's current hub location to the selected target's location. A collective operated at a high level of autonomy by executing actions associated with potential targets independently. The operator was able to influence the collective's actions in order to aid better decision-making, effectively lowering the level of autonomy. Communication from the operator with the collective's entities occurred inside the hub in order to simulate limited real-world communication capabilities. The control mechanisms, for influencing the collective, were communicated to the specified hub. Two visualizations were designed and evaluated in order to determine which visualization provided better transparency by facilitating the operator's perception of the collectives' states, comprehension of the collectives' decision-making processes, and means to influence future collectives' actions. 

\begin{figure}[bp!]
\centering
	\includegraphics[width=\textwidth]{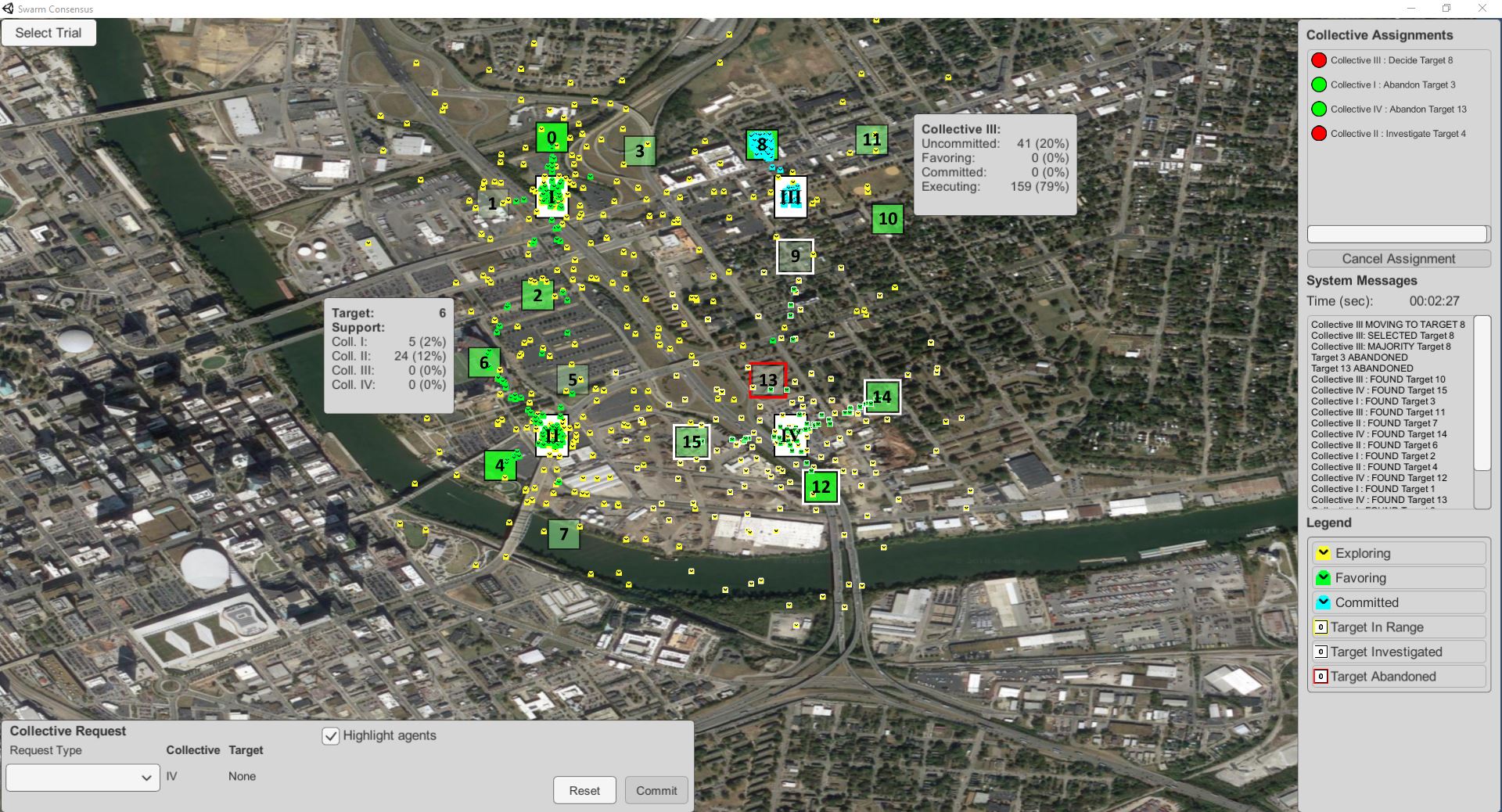}
	\caption{The Individual Agents (IA) interface two and half minutes into a trail, showing four collectives (rectangles with Roman numerals), and the sixteen discovered targets (rectangles with integers). The target's value is represented by the green color, where higher values were brighter. The legend in the lower right corner identifies the individual collective entity state information and target range information.}
    \label{fig: IA Interface}
\end{figure}

\subsection{Individual Agents Interface}
\label{sec: IA Interface}

The Individual Agents (IA) interface, see Figure \ref{fig: IA Interface}, exemplifies a traditional collective visualization by displaying the location of all the individual collective entities \citep{Roundtree20191}. The interface was divided into three primary areas: 1) the central map, 2) the collective request area, and 3) the monitor area. The map, located at the center of the interface visualizes the respective hubs, their individual entities, discovered targets, and other associated information. Both the collectives and targets were rectangular boxes with distinguishing identifiers located at the center of the icon. The collectives had Roman numeral identifiers (I-IV), while the targets used integers (0-15). Discovered targets initially were white and transitioned to a green color when at least two individual collective entities evaluated the target. The highest valued targets were a bright opaque green (e.g., Target 0 in Figure \ref{fig: IA Interface}), while lower valued targets had a more translucent green color (e.g., Target 9 in Figure \ref{fig: IA Interface}). Targets that were within the collective's 500 $m$ search range had different colored outlines, depending on the collective's state: explored targets that were not currently favored had yellow outlines, explored targets that were favored had white outlines (e.g., Target 12 in Figure \ref{fig: IA Interface}), and targets that were abandoned have red outlines (e.g., Target 13 in Figure \ref{fig: IA Interface}).

The individual collective entities began each trial by exploring the environment in an uncommitted state, which transitioned to favoring as targets were assessed and supported. The individual collective entities committed to a target once 30\% of the collective (60 individual entities) favored a particular target. The collective executed a move to the selected target's location once 50\% of the collective (100 entities) favored the target. The individual collective entities' state information was conveyed via individual collective entity color coding: uncommitted (yellow), favoring (green), committed (blue), and executing (blue). A legend of the collective entities' and target border colors was provided in the lower right-hand corner, see Figure \ref{fig: IA Interface}. The number of individual collective entities in a particular state, or supporting a target was provided via the collective hub and target information pop-up displays, which provided detailed information, represented as gray rectangular boxes, displayed directly on the map in Figure \ref{fig: IA Interface}. The information displays, when accessed, appeared in a particular location relative to the respective collective's hub or target. The operator was able to move the information displays by dragging the pop-up display to a desired location.

The operator had the ability to influence an individual collectives' current state via the collective request area, located on the lower left-hand side of Figure \ref{fig: IA Interface}. The \textit{investigate} command permitted increasing a collective's support for an operator specific target. Ten uncommitted entities (5\% of the collective population) transitioned to the favoring state after receiving and acknowledging the investigate command. Additional support for the same target was achieved by reissuing the investigate command repeatedly. The abandon command reduced a collective's support for a specific target by transitioning favoring individual entities to the uncommitted state. The \textit{abandon} command only needed to be issued once in order for the collective to ignore a specified target. A collective's entities stopped exploring alternative targets and moved to the operator selected target when the \textit{decide} command was issued, which was a valid request when at least 30\% of the collective supported the operator specified target. 

The collective assignments area logged the operator's issued commands, shown in the upper right-hand corner of the monitor area in Figure \ref{fig: IA Interface}. The log displayed what commands were issued with respect to particular collectives, for example, Collective I: Abandon Target 3. The green and red circles next to each command signified whether the command was completed (red) or currently active (green). An investigate command initially had a green circle and transitioned to red once ten individual entities received and acknowledged the investigate command for a particular target. Issued abandon commands for a particular collective and target remained active. Once a collective reached a decision, all prior commands associated with that particular collective were removed from the collective assignments log. The only command the operator was able to cancel was the abandon command, which required selecting the desired abandon command, in Figure \ref{fig: IA Interface}, the ``Collective I: Abandon Target 3'', selecting the cancel assignment button.

System messages indicated the actions taken by the operator and collectives. The illegal message was displayed when an operator requested an invalid command, and explained why the requested action was not viable. Three situations resulted in illegal messages. The first arose when the operator attempted to issue an investigate command for targets that were outside of the collective's search region. The second situation occurred when the operator attempted to abandon newly discovered targets that did not have an assigned value (white targets). The last message arouse when operators attempted to issue decide commands when less than 30\% of the collective supported a target. 

\subsection{Collective Interface}

\begin{figure}[h]
\centering
	\includegraphics[width=\textwidth]{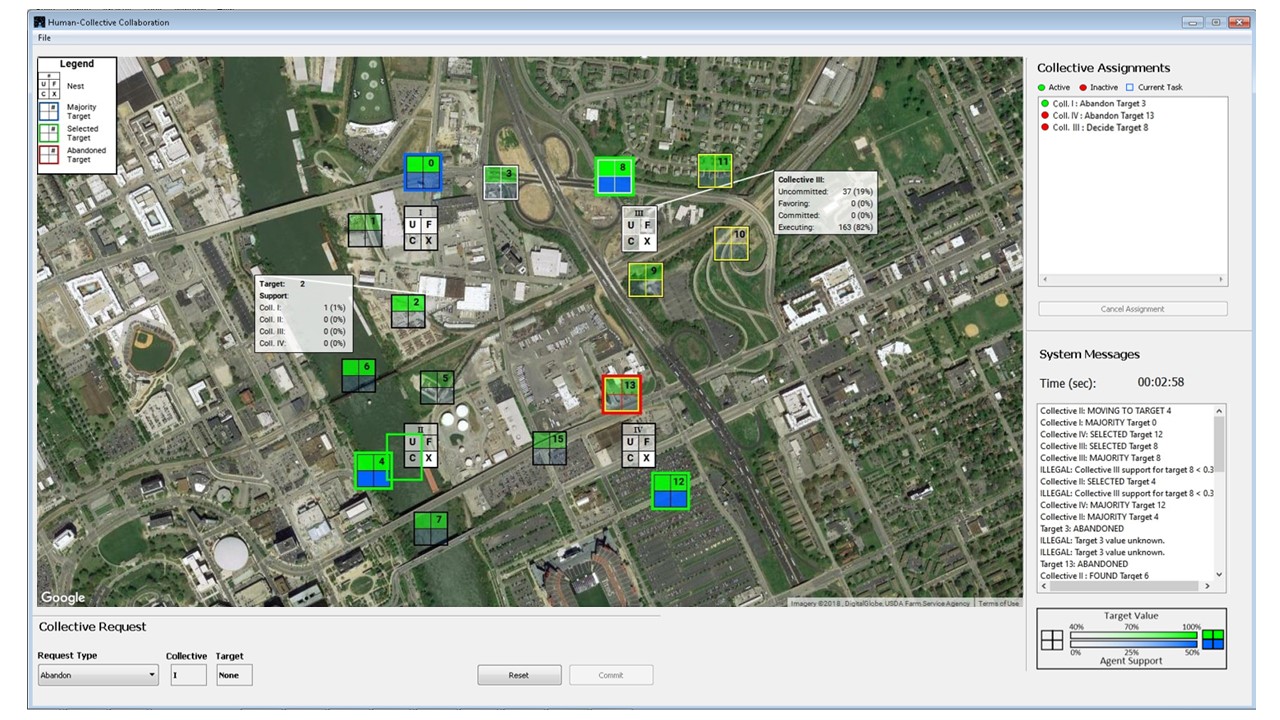}
	\caption{The Collective interface mid-way through a trail scenario, showing the current locations of the four collectives (rectangles with Roman numerals) and the locations of the discovered targets (green and blue squares with integer identifiers). The top half of each target indicates the target's relative value (green) and the bottom half indicates the support of the highest supporting collective (blue). The legend in the upper left hand corner identifies the target range information.}
    \label{fig: Collective Interface}
\end{figure}

The Collective interface, shown in Figure \ref{fig: Collective Interface}, provides an abstract visualization that does not represent individual collectives' entities. The Collective interface was divided into the same three primary areas as the IA interface. The collectives were represented as gray and white rectangles with four quadrants and Roman numeral identifiers located at the top center of the icon. Collective state information was conveyed via the collective icon's quadrants, color coding, and information windows. The collective icon's contained four state quadrants (uncommitted (U), favoring (F), committed (C), and executing (X)), which represented the number of individual entities currently in each state, where a brighter white quadrant equated to larger numbers of individual collective entities. The square target icons had integer identifiers positioned on the upper right hand corner. Targets contained two sections, where the top green section represented the target's value, where the brighter and more opaque the green, the higher the value (e.g., Target 8 in Figure \ref{fig: Collective Interface}). The blue section indicated the number of individual entities favoring a particular target, where the brighter and more opaque the blue, the higher the number of collective entities (e.g., Target 12 in Figure \ref{fig: Collective Interface}). 

The collective interface operated similarly to the IA interface with some distinctions. A target was outlined in blue, demonstrated by Target 0 in Figure \ref{fig: Collective Interface}, when the collective's support exceeded 30\%. The target transitioned to a green outline and the collective was outlined in green when the collective began executing a move to the target's location. The collective's outline moved from the hub to the target's location to indicate the hub's transition to the selected target. The interface's legend appeared in the upper left corner, see Figure \ref{fig: Collective Interface}.

\section{Experimental Design}

The primary research question for the between-evaluation analysis was to determine \textit{which visualization achieved better transparency?} Four secondary questions were developed in order to investigate how the visualization impacted a direct transparency factor, exclusive of trust. The first research question ($R_{1}$) focused on understanding \textit{how the visualization influenced the operator}. Individual differences, such as experience level, will impact an operator's ability to interact with the visualization effectively and cause different responses (i.e., loss of situational awareness or more workload). A visualization that can aid operators with difference capabilities is desired. The explainability factor was encompassed as $R_{2}$, which explored whether \textit{the visualization promoted operator comprehension}. Perception and comprehension of the visualized information are necessary to inform future actions. Understanding \textit{which visualization promoted better usability}, $R_{3}$, will aid designers in promoting effective transparency in human-collective systems. The final research question, $R_{4}$, assessed \textit{which visualization promoted better human-collective performance}. A system that performs a task quickly, safely, and successfully is ideal. 

The independent variables included the between visualization variable, IA versus Collective, and trial difficulty (overall, easy, and hard). Trials that had a larger number of high valued targets in closer proximity to a collective's hub were deemed \textit{easy}, while \textit{hard} trials placed high valued targets further away from the hub. The dependent variable details are embedded into the sections associated with each research question. 

\subsection{Experimental Procedure}

The experimental procedure for both evaluations began with a demographic questionnaire, followed by the Mental Rotations test \cite{Vandenberg1978}, after which the participants received training and practiced using the interface. Two practice sessions occurred prior to each trial in order to ensure familiarity with the underlying sequential best-of-\textit{n} ($M_{2}$) and baseline $M_{3}$ models. The $M_{2}$ model trial was always completed first, in order to alleviate any learning effects from using $M_{3}$. The participants were instructed that the objective was to aid each collective in selecting and moving to the highest valued target two sequential times. A trial began once the respective practice session was completed. Each trial was divided into two components (one easy and one hard) of approximately ten minutes each. Splitting each trial into two components allowed the simulation environment to reset with 16 new (not initially visible) targets. The easy trial contained higher valued targets close to the hub, while the hard trial placed high valued targets further away. The easy and hard trial orderings were randomly assigned, and counterbalanced across the participants. The situational awareness (SA) probe questions \cite{Cody2018}, intended to serve as a secondary task, were asked beginning at 50 seconds into the trial and were repeated at one-minute increments. Six SA probes were asked during each trial. The trial was terminated once the team completed eight decisions, two per collective, or once six decisions were made, if the trial length exceeded the ten-minute limit. Decision times were not limited. A post-trial questionnaire was completed after each trial and the post-experiment questionnaire was completed before the evaluation termination. 

\subsection{Participants}

The participants from both evaluations completed a demographic questionnaire, which collected information regarding age, gender, education level, \textit{weekly hours on a desktop or laptop} (0, less than 3, 3-8, and more than 8), and  their \textit{video game proficiency} from little to no proficiency (1) to high proficiency (7). The \textit{Mental Rotation Assessment} \cite{Vandenberg1978} required participants to judge three-dimensional object orientation to assess spatial reasoning within a scoring range of 0 (low) to 24 (high). The mode is reported in parenthesis for questions that required selection to a group.

\subsubsection{IA Evaluation}

Fourteen females and nineteen males (33 total) completed the IA evaluation at Oregon State University. The predominant (25) age range was between 18 to 30 years, with seven participants in the 31 and 50 range and one participant being 60 and older. Many participants were in the process of obtaining (8) or had an undergraduate degree (13), a master's degree (9), or a doctorate degree (1). The mean number of weekly hours participant's used a desktop or laptop was 3.79, with a standard deviation (SD) = 0.5, median = 4, minimum (min) = 2, and maximum (max) = 4. Participants ranked video game proficiency as mean 4.61 (SD = 1.93, median = 5, min = 1, max = 7). The Mental Rotation Assessment \cite{Vandenberg1978} mean was 12.36 (SD = 5.85, median = 12, min = 3, and max = 24) \cite{Roundtree20191}. 

\subsubsection{Collective Evaluation}

Twenty-eight participants, 15 females and 13 males, from Vanderbilt University, completed the Collective evaluation. The majority of participants (24) were between 18 and 30 years old, with four between 31 and 50. Most of the participants completed high school and were in the process of obtaining (11) or had completed an undergraduate degree (13). The weekly hours participant's used a desktop or laptop was slightly higher than that of the IA (mean = 3.86, SD = 0.45, median = 4, min = 2, and max = 4). The participants' video game proficiency was ranked lower than the IA (mean = 3.61, SD = 2.23, median = 2.5, min = 1, and max = 7). The participants' Mental Rotations Assessment scores were also slightly lower (mean = 10.93, SD = 5.58, median = 10, min = 1, and max = 24) \cite{Roundtree20191}.

\subsection{Analysis}

Five participants were excluded from the IA evaluation due to inconsistent methodology (1) and software failure (4). The between visualization analysis is based on 56 participants from both evaluations. The first twelve decisions made per participant using the $M_{2}$ model were analyzed. The majority of the objective metrics were analyzed by SA level (overall ($SA_{O}$), perception ($SA_{1}$), comprehension ($SA_{2}$), and projection ($SA_{3}$)), decision difficulty (overall, easy, and hard), timing with respect to a SA probe question (15 seconds before asking, while asking, or during response to a SA probe question), or per participant. Non-parametric statistical methods, including Mann-Whitney-Wilcoxon tests and Spearman Correlations, were calculated due to a lack of normality. The correlations were with respect to SA Probe Accuracy and Selection Success Rate. The Collective evaluation data was reanalyzed using the same methods. Secondary research question's hypotheses, associated metrics, results, and discussion are presented in Sections 5-8.  

\section{$R_{1}$: Visualization Influence on Human Operator}

Understanding \textit{how the visualization influenced the operator}, $R_{1}$, is necessary to determine if the transparency embedded into the system design aided operators with different capabilities. The associated objective dependent variables were (1) the operator's ability to influence the collective in order to choose the highest \textit{target value}, (2) situational awareness (SA), (3) interface clutter, and (4) the operator's spatial reasoning capability (Mental Rotations Assessment). The relationship between the variables and the corresponding hypotheses, and the direct and indirect transparency factors, are identified in Table \ref{table:Impacts,Variables}. Additional relationships (not identified in Figure \ref{fig: Concept Map}) between the variable and the direct or indirect transparency factors are identified due to correlation analyses.

\begin{table}[h]
\centering
\caption{Visualization influence on the human operator variables, relationship to the hypotheses, and the associated direct and indirect transparency factors, are presented in Figure \ref{fig: Concept Map}.}
\label{table:Impacts,Variables}
\begin{tabular}{?l|c?c|c?c|c|c|c|c|c|c?}
\Cline{1pt}{3-11}
\multicolumn{1}{c}{} & & \multicolumn{9}{c?}{\textbf{Transparency Factors}} \\ \Cline{1pt}{3-11}
\multicolumn{1}{c}{} & & \multicolumn{2}{c?}{\textbf{Direct}} & \multicolumn{7}{c?}{\textbf{Indirect}} \\ \Cline{1pt}{3-11}
\multicolumn{1}{c}{} & & {\multirow[b]{6}{*}{\rotatebox{90}{\textbf{Performance}}}} & {\multirow[b]{6}{*}{\rotatebox{90}{\textbf{Usability}}}} & {\multirow[b]{6}{*}{\rotatebox{90}{\textbf{Capability}}}} & {\multirow[b]{6}{*}{\rotatebox{90}{\textbf{Effectiveness}}}} &  {\multirow[b]{6}{*}{\rotatebox{90}{\textbf{SA}}}} & {\multirow[b]{6}{*}{\rotatebox{90}{\textbf{Satisfaction}}}} & {\multirow[b]{6}{*}{\rotatebox{90}{\textbf{Timing}}}} & {\multirow[b]{6}{*}{\rotatebox{90}{\textbf{Understanding}}}} &  {\multirow[b]{6}{*}{\rotatebox{90}{\textbf{Workload}}}} \\ 
\multicolumn{1}{c}{} & & & & & & & & & & \\ 
\multicolumn{1}{c}{} & & & & & & & & & & \\ 
\multicolumn{1}{c}{} & & & & & & & & & & \\ 
\multicolumn{1}{c}{} & & & & & & & & & & \\ \Cline{1pt}{1-2}
\multicolumn{1}{?c|}{\textbf{Objective Variables}} & {\textbf{Hypotheses}} & & & & & & & & & \\ \Cline{1pt}{1-11}
{Target Value} & $H_{1}$ & & {\checkmark} & & & & & & & \\ \hline
{SA Probe Accuracy} & $H_{1}$ & {\checkmark} & & & \checkmark & {\checkmark} & & & \checkmark & \\ \hline
{Local Clutter Percentage} & $H_{1}$ & {\checkmark} & {\checkmark} & & {\checkmark} & {\checkmark} & & & \checkmark & \\ \hline
{Global Clutter Percentage} & $H_{1}$ & {\checkmark} & {\checkmark} & & {\checkmark} & {\checkmark} & & & \checkmark & \\ \hline
{Mental Rotations Assessment} & $H_{2}$ & {\checkmark} & & {\checkmark} & & & & & & \\ \Cline{1pt}{1-11}
\multicolumn{1}{?c}{\textbf{Subjective Variables}} & \multicolumn{10}{c?}{\textbf{}} \\ \Cline{1pt}{1-11}
{Weekly Hours on a Desktop or Laptop} & $H_{2}$ & {\checkmark} & & {\checkmark} & & & & & & \\ \hline
{Video Game Proficiency} & $H_{2}$ & {\checkmark} & & {\checkmark} & & & & & & \\ \hline
{NASA Task Load Index} & $H_{1}$, $H_{3}$ & {\checkmark} & & {\checkmark} & & & \checkmark & {\checkmark} & & \checkmark \\ \hline
{3-D Situational Awareness Rating Technique} & $H_{1}$ & & & {\checkmark} & \checkmark & {\checkmark} & & & \checkmark & \\ \Cline{1pt}{1-11}
\end{tabular}
\end{table}

Operators may have performed differently depending on their individual differences. It was hypothesized ($H_{1}$) that operators using the Collective visualization will experience significantly higher SA and lower workload. SA represents an operator's ability to perceive and comprehend information in order to project future actions that must be taken in order to fulfill a task \citep{Endsley1995}. Usability influences the perception of information \citep{Roundtree2019} and will impact workload, which is the amount of stress an operator experiences in order to accomplish a task during a particular duration of time \cite{Wickens2004}. It was hypothesized ($H_{2}$) that operators with different individual capabilities will not perform significantly different using the Collective visualization. An ideal visualization will enable operators with different capabilities to perceive, comprehend, and influence collectives relatively the same. Training can alleviate any disparities between operators, but is only intended to supplement the system's design. The operator's attitude and sentiments towards a system, which is dependent on system usability, provides essential information related to the system's design \cite{Kizilcec2016}. Good designs promote higher operator satisfaction. It was hypothesized ($H_{3}$) that operators using the Collective visualization will experience significantly less frustration (i.e., higher satisfaction).

\subsection{Metrics and Results}
\label{section:R1 metrics}

Assessing variables, such as the selected target value for each human-collective decision, is necessary in order to determine whether operators were able to perceive the target value correctly and influence the collectives positively. The objective of the human-collective team was to select the highest valued target for each decision from a range of target values (67 to 100). The selected target value is the average of all target's respective values that were selected by the human-collective teams during a trial. The descriptive statistics for the selected target value per decision difficulty (i.e., overall, easy, and hard) are shown in Table \ref{table:Impacts,Target Value}. Participants using the Collective interface were able to influence the collective to chose higher valued targets, regardless of decision difficulty, on average; however, the Mann-Whitney-Wilcoxon test identified no significant effects between visualizations for the selected target value. 

\begin{table}[h]
\centering
\caption{The selected target value descriptive statistics by decision difficulty, where the maximum possible value was 100 and the minimum possible value was 67.}
\label{table:Impacts,Target Value}
\begin{tabular}{c|c|c|c|}
\cline{2-4}
 & \textbf{Decision Difficulty} & \textbf{Mean (SD)} & \textbf{Median (Min/Max)} \\ \hline
\multicolumn{1}{|c|}{\multirow{3}{*}{IA}} & Overall & 90.29 (7.11) & 95 (67/97) \\ \cline{2-4} 
\multicolumn{1}{|c|}{} & \cellcolor{light-gray}Easy & \cellcolor{light-gray}90.21 (7.29) & \cellcolor{light-gray}95 (67/97) \\ \cline{2-4} 
\multicolumn{1}{|c|}{} & \cellcolor{medium-gray}Hard & \cellcolor{medium-gray}90.4 (6.88) & \cellcolor{medium-gray}94 (68/96) \\ \hline
\multicolumn{1}{|c|}{\multirow{3}{*}{Collective}} & Overall & 92.05 (5.08) & 95 (68/96) \\ \cline{2-4} 
\multicolumn{1}{|c|}{} & \cellcolor{light-gray}Easy & \cellcolor{light-gray}92.09 (5.54) & \cellcolor{light-gray}95 (68/96) \\ \cline{2-4} 
\multicolumn{1}{|c|}{} & \cellcolor{medium-gray}Hard & \cellcolor{medium-gray}92 (4.5) & \cellcolor{medium-gray}95 (78/96) \\ \hline
\end{tabular}
\end{table}

The SA dependent variable was \textit{SA probe accuracy}, which is the percentage of correctly answered SA probes questions used to assess the operator's SA during a trial \cite{Cody2018}. Each question corresponded to the three SA levels: perception, comprehension, and projection \cite{Endsley1995}. Participants were asked five $SA_{1}$, four $SA_{2}$, and three $SA_{3}$ questions. The \textit{$SA_{1}$} questions determined the operator's ability to perceive information about the collectives and targets, such as ``What collectives are investigating Target 3?'' The operator's comprehension of information was represented by the \textit{$SA_{2}$} questions, such as ``Which Collective has achieved a majority support for Target 7?'' \textit{$SA_{3}$} questions related to the operator's ability to estimate the collectives' future state, such as ``Will support for Target 1 decrease?'' An overall SA value, $SA_{O}$, represented the percent of correctly answered SA probes out of 12 total. The SA probe accuracy descriptive statistics are shown in Table \ref{table:Impacts,SA Probe Accuracy} \cite{Roundtree20191}. Operators using the Collective visualization had higher SA probe accuracy, regardless of the SA level. The Mann-Whitney-Wilcoxon tests (n = 56, degrees of freedom (DOF) = 1) found highly significant effects between visualizations for $SA_{O}$ (U = 702, $\rho$ $<$ 0.001) and $SA_{1}$ (U = 714.5, $\rho$ $<$ 0.001). Moderately significant effects were found for $SA_{2}$ (U = 572.5, $\rho$ $<$ 0.01) and $SA_{3}$ (U = 554, $\rho$ $<$ 0.01). 

\begin{table}[h]
\centering
\caption{SA probe accuracy (\%) descriptive statistics by SA level.}
\label{table:Impacts,SA Probe Accuracy}
\begin{tabular}{c|c|c|c|}
\cline{2-4}
 & \textbf{SA Level} & \textbf{Mean (SD)} & \textbf{Median (Min/Max)} \\ \hline
\multicolumn{1}{|c|}{\multirow{4}{*}{IA}} & $SA_{O}$ & 65.3 (18.87) & 68.33 (16.67/83.33) \\ \cline{2-4} 
\multicolumn{1}{|c|}{} & \cellcolor{light-gray}$SA_{1}$ & \cellcolor{light-gray}58.57 (23.05) & \cellcolor{light-gray}60 (20/100) \\ \cline{2-4} 
\multicolumn{1}{|c|}{} & \cellcolor{medium-gray}$SA_{2}$ & \cellcolor{medium-gray}72.32 (21.88) & \cellcolor{medium-gray}75 (25/100) \\ \cline{2-4} \multicolumn{1}{|c|}{} & \cellcolor{dark-gray}$SA_{3}$ & \cellcolor{dark-gray}65.48 (34.52) & \cellcolor{dark-gray}66.67 (0/100) \\ \hline
\multicolumn{1}{|c|}{\multirow{4}{*}{Collective}} & $SA_{O}$ & 89.88 (10.96) & 91.67 (58.33/100) \\ \cline{2-4} 
\multicolumn{1}{|c|}{} & \cellcolor{light-gray}$SA_{1}$ & \cellcolor{light-gray}91.67 (11.11) & \cellcolor{light-gray}100 (66.67/100) \\ \cline{2-4} 
\multicolumn{1}{|c|}{} & \cellcolor{medium-gray}$SA_{2}$ & \cellcolor{medium-gray}88.39 (14.6) & \cellcolor{medium-gray}100 (60/100) \\ \cline{2-4}
\multicolumn{1}{|c|}{} & \cellcolor{dark-gray}$SA_{3}$ & \cellcolor{dark-gray}89.88 (20.46) & \cellcolor{dark-gray}100 (33.33/100) \\ \hline
\end{tabular}
\end{table}

Local and global clutter percentages were analyzed for each SA probe question. Clutter is defined as the area occupied by objects on a display, relative to the total area of the display \cite{Wickens2004}. Presenting too much information in close proximity to one another will require the operator to search longer for information \cite{Wickens2004} and can negatively influence the accuracy of the SA probe questions. Area coverage for each 2-D item was calculated by the number of pixels the item covered on the computer visualization. The conversion between meters and pixels was different for each visualization due to differences in the display monitor size and software program. One meter for the IA visualization was approximately 1.97 pixels per meter and the Collective visualization was approximately 2.3 pixels per meter. The \textit{local clutter percentage} variable was the percentage of area obstructed by items that were displayed within the 500 $m$ (i.e., approximately 254 pixels for the IA visualization and 218 pixels for the Collective visualization) circular radius from the center of the collective, or target of interest in the SA probe question. Collective IV, for example, is the collective of interest in the following SA probe question: ``What is the highest value target available to Collective IV?'' The items obstructing the 500 $m$ radius when using the IA visualization, in Figure \ref{fig: IA Interface}, for the previous SA probe question are: the Collective IV, Targets 9 and 12-15, and 200 individual entities. Some SA probe questions encompassed more than one collective or target of interest, which required calculating the local clutter percentage for each collective or target and summing the values together. All calculations first required converting meters into pixels in order to ensure equivalent units. The Collective visualization computer display size was unknown; therefore, local and global clutter percentage calculations use the corresponding item and computer display dimensions from the IA visualization. Local clutter was calculated using Equation \ref{eq:Local Clutter}:

\begin{table}[bp!]
\centering
\caption{Local clutter percentage descriptive statistics 15 seconds before asking, while asking, and during response to SA probe question by SA level.}
\label{table:Impacts,Local Clutter Percentage}
\begin{tabular}{c|c|c|c|c|}
\cline{2-5}
 & \textbf{Timing} & \textbf{SA Level} & \textbf{Mean (SD)} & \textbf{Median (Min/Max)} \\ \hline
\multicolumn{1}{|c|}{\multirow{12}{*}{IA}} & \multirow{4}{*}{Before} & $SA_{O}$ & 33.6 (21.66) & 26.13 (9/124) \\ \cline{3-5} 
\multicolumn{1}{|c|}{} & & \cellcolor{light-gray}$SA_{1}$ & \cellcolor{light-gray}30.79 (19.53) & \cellcolor{light-gray}24.4 (9/122.33) \\ \cline{3-5} 
\multicolumn{1}{|c|}{} & & \cellcolor{medium-gray}$SA_{2}$ & \cellcolor{medium-gray}41.54 (24.39) & \cellcolor{medium-gray}37.75 (9/124) \\ \cline{3-5} 
\multicolumn{1}{|c|}{} & & \cellcolor{dark-gray}$SA_{3}$ & \cellcolor{dark-gray}28.61 (19.36) & \cellcolor{dark-gray}22.23 (9.08/97.7) \\ \cline{2-5} 
\multicolumn{1}{|c|}{} & \multirow{4}{*}{Asking} & $SA_{O}$ & 34.42 (22.16) & 28 (9/124) \\ \cline{3-5} 
\multicolumn{1}{|c|}{} & & \cellcolor{light-gray}$SA_{1}$ & \cellcolor{light-gray}31.91 (20.54) & \cellcolor{light-gray}25 (9/122) \\ \cline{3-5} 
\multicolumn{1}{|c|}{} & & \cellcolor{medium-gray}$SA_{2}$ & \cellcolor{medium-gray}41.67 (24.74) & \cellcolor{medium-gray}37.17 (9/124) \\ \cline{3-5} 
\multicolumn{1}{|c|}{} & & \cellcolor{dark-gray}$SA_{3}$ & \cellcolor{dark-gray}29.73 (19.54) & \cellcolor{dark-gray}24.21 (9/97.67) \\ \cline{2-5} 
\multicolumn{1}{|c|}{} & \multirow{4}{*}{Responding} & $SA_{O}$ & 34.26 (22.25) & 27.5 (8/124) \\ \cline{3-5} 
\multicolumn{1}{|c|}{} & & \cellcolor{light-gray}$SA_{1}$ & \cellcolor{light-gray}31.84 (20.61) & \cellcolor{light-gray}24.83 (9/122) \\ \cline{3-5} 
\multicolumn{1}{|c|}{} & & \cellcolor{medium-gray}$SA_{2}$ & \cellcolor{medium-gray}41.28 (24.96) & \cellcolor{medium-gray}36.5 (8/124) \\ \cline{3-5} 
\multicolumn{1}{|c|}{} & & \cellcolor{dark-gray}$SA_{3}$ & \cellcolor{dark-gray}29.68 (19.6) & \cellcolor{dark-gray}24 (9/98) \\ \hline
\multicolumn{1}{|c|}{\multirow{12}{*}{Collective}} & \multirow{4}{*}{Before} & $SA_{O}$ & 35.42 (28.08) & 25.44 (9/177) \\ \cline{3-5} 
\multicolumn{1}{|c|}{} & & \cellcolor{light-gray}$SA_{1}$ & \cellcolor{light-gray}34.09 (29.09) & \cellcolor{light-gray}24.04 (9/151.9) \\ \cline{3-5} 
\multicolumn{1}{|c|}{} & & \cellcolor{medium-gray}$SA_{2}$ & \cellcolor{medium-gray}35.81 (27.53) & \cellcolor{medium-gray}26.86 (10.3/177) \\ \cline{3-5} 
\multicolumn{1}{|c|}{} & & \cellcolor{dark-gray}$SA_{3}$ & \cellcolor{dark-gray}37.98 (26.78) & \cellcolor{dark-gray}28.23 (10.38/131.47) \\ \cline{2-5} 
\multicolumn{1}{|c|}{} & \multirow{4}{*}{Asking} & $SA_{O}$ & 35.37 (28.78) & 25.75 (9/176.5) \\ \cline{3-5} 
\multicolumn{1}{|c|}{} & & \cellcolor{light-gray}$SA_{1}$ & \cellcolor{light-gray}34.24 (30.37) & \cellcolor{light-gray}23.63 (9/176.5) \\ \cline{3-5} 
\multicolumn{1}{|c|}{} & & \cellcolor{medium-gray}$SA_{2}$ & \cellcolor{medium-gray}36.47 (26.76) & \cellcolor{medium-gray}27.4 (10.2/130.4) \\ \cline{3-5} 
\multicolumn{1}{|c|}{} & & \cellcolor{dark-gray}$SA_{3}$ & \cellcolor{dark-gray}36.35 (28.27) & \cellcolor{dark-gray}27.25 (9/147.56) \\ \cline{2-5} 
\multicolumn{1}{|c|}{} & \multirow{4}{*}{Responding} & $SA_{O}$ & 35.6 (29.19) & 25.8 (9/176.4) \\ \cline{3-5} 
\multicolumn{1}{|c|}{} & & \cellcolor{light-gray}$SA_{1}$ & \cellcolor{light-gray}34.29 (29.84) & \cellcolor{light-gray}25 (9/176.4) \\ \cline{3-5}
\multicolumn{1}{|c|}{} & & \cellcolor{medium-gray}$SA_{2}$ & \cellcolor{medium-gray}36.55 (27.98) & \cellcolor{medium-gray}27 (10/130) \\ \cline{3-5}
\multicolumn{1}{|c|}{} & & \cellcolor{dark-gray}$SA_{3}$ & \cellcolor{dark-gray}37.24 (29.86) & \cellcolor{dark-gray}26.57 (9/147.57) \\ \hline
\end{tabular}
\end{table}

\begin{equation}
Local Clutter Percentage (\%) = \sum \left(\frac{LHA + LHTA + LTA + LAICE + LTIW + LCIW}{\pi\cdot500^2}\right) \cdot 100, 
\label{eq:Local Clutter}
\end{equation} where LHA represents the area corresponding to the number of collective hubs (2464 $pixels^{2}$ per hub) inside the 500 $m$ radius. The area corresponding to the number of highlighted targets (2350 $pixels^{2}$ per highlighted target), which have outlines and are in range of the currently selected collective are represented as LHTA, while the targets that are not highlighted (1720 $pixels^{2}$ per target) are denoted as LTA. LAICE represents the area corresponding to the number of individual collective entities (64 $pixels^{2}$ per agent) inside of the 500 $m$ radius, and was excluded from the Collective visualization local clutter percentage calculation, because no individual entities were displayed. The individual collective entities were confined to the 500 $m$ search radius around their respective collective hub; therefore, the calculation assumes that the 200 entities associated with each collective are inside of the local radius. The area corresponding to the number of target information pop-up windows (32922 $pixels^{2}$ per target information pop-up window) is represented as LTIW, and the corresponding collective information pop-up windows (25740 $pixels^{2}$ per collective information pop-up window) are represented as LCIW. Only the target or collective information pop-up windows that belong to targets or collectives inside of the 500 $m$ radius are considered.

The local clutter percentage descriptive statistics 15 seconds before asking, while asking, and during a SA probe response are provided in Table \ref{table:Impacts,Local Clutter Percentage}. The maximum local clutter percentage was 177\%, which indicated that the area covered by the associated items of the collective or target of interest in the SA probe exceeded the 500 $m$ radius. Local clutter percentages larger than 100\% were attributed to the area covered by the collective and target information pop-up windows. The location of the information pop-up windows were not recorded; therefore, the maximum area coverage was considered when information pop-up windows did not occlude items in the environment. The maximum area coverage condition was not reflective of the real trial environment, where information pop-up windows covered items on the central map. The IA visualization had lower local clutter percentage, regardless of when the metric was collected for $SA_{O}$, $SA_{1}$, and $SA_{3}$; although, no significant effects were found between visualizations. The Spearman correlation analysis revealed no correlations were found between local clutter percentage and SA probe accuracy. 

The \textit{global clutter percentage}, calculated using Equation \ref{eq:Global Clutter}, was the percentage of area obstructed by all objects displayed on the entire IA computer display (2073600 $pixels^{2}$), since the Collective computer display was unknown.

\begin{table}[bp!]
\centering
\caption{Global clutter percentage descriptive statistics 15 seconds before asking, while asking, and during response to SA probe question by SA level.}
\label{table:Impacts,Global Clutter Percentage}
\begin{tabular}{c|c|c|c|c|}
\cline{2-5}
 & \textbf{Timing} & \textbf{SA Level} & \textbf{Mean (SD)} & \textbf{Median (Min/Max)} \\ \hline
\multicolumn{1}{|c|}{\multirow{12}{*}{IA}} & \multirow{4}{*}{Before} & $SA_{O}$ & 30.2 (3.06) & 28.83 (27/40.22) \\ \cline{3-5} 
\multicolumn{1}{|c|}{} & & \cellcolor{light-gray}$SA_{1}$ & \cellcolor{light-gray}29.88 (2.8) & \cellcolor{light-gray}28.5 (27/40.22) \\ \cline{3-5} 
\multicolumn{1}{|c|}{} & & \cellcolor{medium-gray}$SA_{2}$ & \cellcolor{medium-gray}30.41 (3.05) & \cellcolor{medium-gray}29.1 (27/40) \\ \cline{3-5}
\multicolumn{1}{|c|}{} & & \cellcolor{dark-gray}$SA_{3}$ & \cellcolor{dark-gray}30.45 (3.45) & \cellcolor{dark-gray}28.85 (27/40) \\ \cline{2-5} 
\multicolumn{1}{|c|}{} & \multirow{4}{*}{Asking} & $SA_{O}$ & 30.25 (3.13) & 29 (27/40) \\ \cline{3-5} 
\multicolumn{1}{|c|}{} & & \cellcolor{light-gray}$SA_{1}$ & \cellcolor{light-gray}29.95 (2.91) & \cellcolor{light-gray}28.58 (27/40) \\ \cline{3-5}
\multicolumn{1}{|c|}{} & & \cellcolor{medium-gray}$SA_{2}$ & \cellcolor{medium-gray}30.41 (3.12) & \cellcolor{medium-gray}29 (27/40) \\ \cline{3-5} 
\multicolumn{1}{|c|}{} & & \cellcolor{dark-gray}$SA_{3}$ & \cellcolor{dark-gray}30.52 (3.49) & \cellcolor{dark-gray}29 (27/40) \\ \cline{2-5} 
\multicolumn{1}{|c|}{} & \multirow{4}{*}{Responding} & $SA_{O}$ & 30.09 (3.02) & 29 (27/40) \\ \cline{3-5} 
\multicolumn{1}{|c|}{} & & \cellcolor{light-gray}$SA_{1}$ & \cellcolor{light-gray}29.83 (2.81) & \cellcolor{light-gray}28.5 (27/40) \\ \cline{3-5} 
\multicolumn{1}{|c|}{} & & \cellcolor{medium-gray}$SA_{2}$ & \cellcolor{medium-gray}30.22 (3) & \cellcolor{medium-gray}29 (27/40) \\ \cline{3-5} 
\multicolumn{1}{|c|}{} & & \cellcolor{dark-gray}$SA_{3}$ & \cellcolor{dark-gray}30.37 (3.38) & \cellcolor{dark-gray}28.79 (27/40) \\ \hline
\multicolumn{1}{|c|}{\multirow{12}{*}{Collective}} & \multirow{4}{*}{Before} & $SA_{O}$ & 31.37 (4.97) & 29.21 (27.88/53) \\ \cline{3-5} 
\multicolumn{1}{|c|}{} & & \cellcolor{light-gray}$SA_{1}$ & \cellcolor{light-gray}31.38 (5) & \cellcolor{light-gray}29.13 (28/52.4) \\ \cline{3-5} 
\multicolumn{1}{|c|}{} & & \cellcolor{medium-gray}$SA_{2}$ & \cellcolor{medium-gray}31.25 (5.09) & \cellcolor{medium-gray}29.21 (27.88/53) \\ \cline{3-5} 
\multicolumn{1}{|c|}{} & & \cellcolor{dark-gray}$SA_{3}$ & \cellcolor{dark-gray}31.56 (4.76) & \cellcolor{dark-gray}29.54 (28/52) \\ \cline{2-5} 
\multicolumn{1}{|c|}{} & \multirow{4}{*}{Asking} & $SA_{O}$ & 31.43 (5.13) & 29.17 (28/53) \\ \cline{3-5} 
\multicolumn{1}{|c|}{} & & \cellcolor{light-gray}$SA_{1}$ & \cellcolor{light-gray}31.24 (5.26) & \cellcolor{light-gray}29 (28/53) \\ \cline{3-5} 
\multicolumn{1}{|c|}{} & & \cellcolor{medium-gray}$SA_{2}$ & \cellcolor{medium-gray}31.52 (5.2) & \cellcolor{medium-gray}29.22 (28/51.8) \\ \cline{3-5} 
\multicolumn{1}{|c|}{} & & \cellcolor{dark-gray}$SA_{3}$ & \cellcolor{dark-gray}31.69 (4.78) & \cellcolor{dark-gray}29.75 (28/48) \\ \cline{2-5} 
\multicolumn{1}{|c|}{} & \multirow{4}{*}{Responding} & $SA_{O}$ & 31.41 (5.15) & 29 (28/53) \\ \cline{3-5} 
\multicolumn{1}{|c|}{} & & \cellcolor{light-gray}$SA_{1}$ & \cellcolor{light-gray}31.43 (5.43) & \cellcolor{light-gray}29 (28/53) \\ \cline{3-5} 
\multicolumn{1}{|c|}{} & & \cellcolor{medium-gray}$SA_{2}$ & \cellcolor{medium-gray}31.34 (5.08) & \cellcolor{medium-gray}29 (28/51.5) \\ \cline{3-5}
\multicolumn{1}{|c|}{} & & \cellcolor{dark-gray}$SA_{3}$ & \cellcolor{dark-gray}31.49 (4.66) & \cellcolor{dark-gray}29.31 (28/48) \\ \hline
\end{tabular}
\end{table}

\begin{equation}
Global Clutter Percentage = \left(\frac{ICA + GHA + GHTA + GTA + GAICE + GTIW + GCIW}{2073600}\right) \cdot 100, 
\label{eq:Global Clutter}
\end{equation} where ICA represents the area of the static interface components (493414 $pixels^{2}$), which encompass the program bar, the Microsoft Windows program bar, the select trial button, the collective request area, and the monitor area. GHA represents the area covered by Collectives I-IV (9856 $pixels^{2}$), which were visible throughout the duration of a trial. The area corresponding to the number of highlighted targets (2350 $pixels^{2}$ per highlighted target), which have outlines and are in range of the currently selected collective are represented as GHTA. Remaining targets that are not highlighted (1720 $pixels^{2}$ per target), are represented as GTA. GAICE represents the area encompassed by 800 individual collective entities (51200 $pixels^{2}$), which was only considered for the IA visualization. The area corresponding to the number of target information pop-up windows (32922 $pixels^{2}$ per target information pop-up window) is represented as GTIW and the corresponding collective information pop-up windows is represented as GCIW (25740 $pixels^{2}$ per collective information pop-up window).

The global clutter percentage descriptive statistics 15 seconds before asking, while asking, and during response to a SA probe question are shown in Table \ref{table:Impacts,Global Clutter Percentage}. The IA visualization had a lower global clutter percentage, regardless of when the metric was assessed across all SA levels. The Mann-Whitney-Wilcoxon tests found highly significant effects between visualizations when responding to a SA probe question (n = 670, DOF = 1) for $SA_{O}$ (U = 64442, $\rho$ $<$ 0.001). Moderate significant effects were found for $SA_{O}$ 15 seconds before asking a SA probe (U = 64188, $\rho$ $<$ 0.01) and while asking a SA probe question (U = 63728, $\rho$ $<$ 0.01). Significant effects were found 15 seconds before asking a SA probe question for $SA_{1}$ (n = 294, DOF = 1, U = 12487, $\rho$ = 0.02) and $SA_{3}$ (n = 152, DOF = 1, U = 3445.5, $\rho$ = 0.03); while asking a SA probe question for $SA_{1}$ (n = 294, DOF = 1, U = 12301, $\rho$ = 0.03) and $SA_{3}$ (n = 152, DOF = 1, U = 3452, $\rho$ = 0.05); and during the response to a SA probe question for $SA_{1}$ (n = 294, DOF = 1, U = 12216, $\rho$ = 0.04). The Spearman correlation analysis revealed a weak correlation for the Collective visualization for $SA_{1}$ between global clutter percentage 15 seconds before asking a SA probe question and SA probe accuracy (r = 0.16, $\rho$ = 0.05).

The Mental Rotations Assessment \cite{Vandenberg1978}, which assessed the operator's spatial reasoning, identified no significant effects between visualizations. A Spearman correlation analysis revealed weak correlations between the Mental Rotations Assessment and SA probe accuracy when using the IA visualization for $SA_{O}$ (r = 0.17, $\rho$ $<$ 0.01), $SA_{1}$ (r = 0.18, $\rho$ = 0.03), and $SA_{2}$ (r = 0.27, $\rho$ $<$ 0.01). The Mann-Whitney-Wilcoxon tests identified no significant effects between visualizations for the weekly hours spent using a desktop or laptop and video game proficiency. Weak correlations were found between weekly hours using a desktop or laptop and SA probe accuracy for the IA visualization for $SA_{O}$ (r = 0.12, $\rho$ = 0.04) and $SA_{1}$ (r = 0.21, $\rho$ = 0.01), as well as when using the Collective visualization for $SA_{2}$ (r = 0.21, $\rho$ = 0.02). No correlations were found between video game proficiency and SA probe accuracy. 

The NASA Task Load Index (\textit{NASA-TLX}) assessed the six workload subscales and the weighted overall workload \cite{Hart1988}. The descriptive statistics for the NASA-TLX demands imposed on the operator are presented in Table \ref{table:Impacts,NASA TLX}. The Collective visualization imposed a lower overall workload, had lower physical and temporal demands, and caused less frustration. The IA visualization imposed a lower mental demand, which had a significant effect between visualizations (n = 56, DOF = 1, U = 515, $\rho$ = 0.04) and less effort. The IA visualization had higher performance with a highly significant effect between visualizations (n = 56, DOF = 1, U = 159.5, $\rho$ $<$ 0.001).

\begin{table}[h]
\centering
\caption{NASA-TLX descriptive statistics.}
\label{table:Impacts,NASA TLX}
\begin{tabular}{c|c|c|c|}
\cline{2-4}
 & \textbf{Overall and Subscales} & \textbf{Mean (SD)} & \textbf{Median (Min/Max)} \\ \hline
\multicolumn{1}{|c|}{\multirow{7}{*}{IA}} & Overall & 62.14 (14.81) & 65.67 (24/85.67) \\ \cline{2-4} 
\multicolumn{1}{|c|}{} & \cellcolor{gray1}Mental & \cellcolor{gray1}19.25 (8.8) & \cellcolor{gray1}20 (0/33.33) \\ \cline{2-4} 
\multicolumn{1}{|c|}{} & \cellcolor{gray2}Physical & \cellcolor{gray2}1.68 (3.32) & \cellcolor{gray2}0 (0/13) \\ \cline{2-4} 
\multicolumn{1}{|c|}{} & \cellcolor{gray3}Temporal & \cellcolor{gray3}11.75 (8.24) & \cellcolor{gray3}9.67 (0/28.33) \\ \cline{2-4} 
\multicolumn{1}{|c|}{} & \cellcolor{gray4}Performance & \cellcolor{gray4}10.69 (5.87) & \cellcolor{gray4}8.83 (2.67/25) \\ \cline{2-4} 
\multicolumn{1}{|c|}{} & \cellcolor{gray5}Effort & \cellcolor{gray5}11.35 (6.68) & \cellcolor{gray5}11 (2.67/28.33) \\ \cline{2-4} 
\multicolumn{1}{|c|}{} & \cellcolor{gray6}Frustration & \cellcolor{gray6}7.43 (8.36) & \cellcolor{gray6}4.67 (0/33.33) \\ \hline
\multicolumn{1}{|c|}{\multirow{7}{*}{Collective}} & Overall & 57.06 (16.47) & 56.83 (5.67/83.33) \\ \cline{2-4} 
\multicolumn{1}{|c|}{} & \cellcolor{gray1}Mental & \cellcolor{gray1}23.58 (6.34) & \cellcolor{gray1}25 (3/31.67) \\ \cline{2-4} 
\multicolumn{1}{|c|}{} & \cellcolor{gray2}Physical & \cellcolor{gray2}0.46 (1.17) & \cellcolor{gray2}0 (0/4.67) \\ \cline{2-4} 
\multicolumn{1}{|c|}{} & \cellcolor{gray3}Temporal & \cellcolor{gray3}10.94 (7.67) & \cellcolor{gray3}10.33 (0/24) \\ \cline{2-4} 
\multicolumn{1}{|c|}{} & \cellcolor{gray4}Performance & \cellcolor{gray4}5.1 (4.7) & \cellcolor{gray4}3.67 (0/21.33) \\ \cline{2-4} 
\multicolumn{1}{|c|}{} & \cellcolor{gray5}Effort & \cellcolor{gray5}12.32 (6.26) & \cellcolor{gray5}13 (2/25.33) \\ \cline{2-4} 
\multicolumn{1}{|c|}{} & \cellcolor{gray6}Frustration & \cellcolor{gray6}4.65 (6.84) & \cellcolor{gray6}1.83 (0/30) \\ \hline
\end{tabular}
\end{table}

The 3-D Situational Awareness Rating Technique (\textit{SART}) measured the operator's perceived situational understanding, demand on attentional resources, and supply of attentional resources \cite{Selcon1991}. An overall score was calculated using the standard calculation. The SART descriptive statistics are presented in Table \ref{table:Impacts,SART} \cite{Roundtree20191}. The minimum SART score was -1, which was unexpected as a negative score requires the supply of attentional resources to exceed the demand on attentional resources and a low perceived situational understanding. Both of these conditions are highly unlikely. The Collective visualization had a higher overall score, more situational understanding, high demands of attentional resources (although nearly the same as the IA visualization), and more supply of attentional resources, compared to the IA visualization. The Mann-Whitney-Wilcoxon test indicated moderately significant effects between visualizations for the overall score (n = 56, DOF = 1, U = 560, $\rho$ $<$ 0.01), situational understanding (n = 56, DOF = 1, U = 561, $\rho$ $<$ 0.01), and supply of attentional resources (n = 56, DOF = 1, U = 561, $\rho$ $<$ 0.01). 

\begin{table}[h]
\centering
\caption{SART descriptive statistics (1-low, 7-high).}
\label{table:Impacts,SART}
\begin{tabular}{c|c|c|c|}
\cline{2-4}
 & \textbf{Overall and Dimensions} & \textbf{Mean (SD)} & \textbf{Median (Min/Max)} \\ \hline
\multicolumn{1}{|c|}{\multirow{4}{*}{IA}} & Overall & 4.64 (2.6) & 4.5 (-1/10) \\ \cline{2-4} 
\multicolumn{1}{|c|}{} & \cellcolor{light-gray}Situational Understanding & \cellcolor{light-gray}4.96 (1.53) & \cellcolor{light-gray}5 (2/7) \\ \cline{2-4} 
\multicolumn{1}{|c|}{} & \cellcolor{medium-gray}Demands on Attentional Resources & \cellcolor{medium-gray}5.04 (1.2) & \cellcolor{medium-gray}5 (2/7) \\ \cline{2-4} 
\multicolumn{1}{|c|}{} & \cellcolor{dark-gray}Supply of Attentional Resources & \cellcolor{dark-gray}4.71 (1.36) & \cellcolor{dark-gray}5 (1/7) \\ \hline
\multicolumn{1}{|c|}{\multirow{4}{*}{Collective}} & Overall & 6.68 (2.26) & 6.5 (3/13) \\ \cline{2-4} 
\multicolumn{1}{|c|}{} & \cellcolor{light-gray}Situational Understanding & \cellcolor{light-gray}6.07 (0.9) & \cellcolor{light-gray}6 (4/7) \\ \cline{2-4} 
\multicolumn{1}{|c|}{} & \cellcolor{medium-gray}Demands on Attentional Resources & \cellcolor{medium-gray}5.07 (1.18) & \cellcolor{medium-gray}5 (1/6) \\ \cline{2-4}
\multicolumn{1}{|c|}{} & \cellcolor{dark-gray}Supply of Attentional Resources & \cellcolor{dark-gray}5.68 (1.09) & \cellcolor{dark-gray}6 (3/7) \\ \hline
\end{tabular}
\end{table}

\subsection{Discussion}

The analysis of how visualization influenced operators suggests that the Collective visualization promoted better transparency. The variables that directly supported $H_{1}$ are the SA probe accuracy and SART. $H_{1}$ was supported, because operators using the Collective visualization had significantly higher objective and subjective SA and lower overall workload. Transparency embedded into the Collective visualization, via color-coded icons (i.e., target value) and outlines, state information identified on the collective icon, information provided in the collective and target information pop-up windows, and feedback provided in the Collective Assignments and System Messages areas, promoted better perception, comprehension, and projection of future operator actions, making the overall human-collective team more effective. The Collective visualization operators; however, had more local and global clutter, which was mainly attributed to the number of collective and target information pop-up windows that were visible. The increased clutter has both positive and negative implications for transparency. From a system design perspective, clutter is not ideal if operators are unable to perform their tasks effectively. The Collective operators, in this evaluation, who had more global clutter were able to answer more SA probe questions accurately, which suggests that operators were not hindered by the clutter and were able to perform better than their counterparts. The dependence on having the collective and target information pop-up windows visible suggests that the collective state information provided on the collective icon was not as effective as the information pop-up window and there is a need to provide support information on the target icons. Other design strategies must be investigated to improve the efficacy of the collective and target icons for the Collective visualization. Further analyses are also required to determine what contributed to more mental demand, more effort, and less perceived performance using the Collective visualization and whether the additional stress may have been experienced due to positive aspects, such as operators being highly motivated to complete their tasks. The Collective visualization may have required more effort on behalf of the operator in order to understand what the collective was doing compared to the IA visualization that showed the dynamic behavior emerging. Collective visualization operators may have additionally been distracted by the secondary SA probe question task and required more time to enter back into the loop to determine the collective behaviors. 

The Mental Rotation Assessment and video-game proficiency supported $H_{2}$, since operators with different individual capabilities did not perform significantly different using the Collective visualization. One exception to the hypotheses was that more experienced operators may have performed better because of their extensive use of computers, which may have led to faster and more accurate interpretations of information \citep{Wickens2004} (i.e., different types of iconography), or easier access to the supplemental information. Since the exception was observed in both evaluations, the behavior is inherent to working with a computer interface, rather than a particular visualization. Using an abstract collective visualization will mitigate the need for particular operator capabilities to perform the sequential best-of-\textit{n} decision-making task. Operators using the Collective visualization experienced less frustration, which supports $H_{3}$. Dissatisfaction (i.e., frustration) transpires when the system is not transparent and prohibits the operator from understanding what is currently happening, or there is too much clutter and the interface appears noisy \citep{Preece2007}. The abstract visualization may be a solution to mitigate dissatisfaction.  

\section{$R_{2}$: Visualization Promotion of Human Operator Comprehension}

The explainability direct transparency factor was encompassed in $R_{2}$, which was interested in determining whether \textit{the visualization promoted operator comprehension}, by embedding transparency into the system design. Perception and comprehension of the presented information are necessary to inform future operator actions. The associated objective dependent variables were (1) SA, (2) collective and target left- or right-clicks, (3) the percentage of times the highest value target was abandoned, and (4) whether the information pop-up window was open when a target was abandoned. The relationship between the variables and the corresponding hypotheses, and the direct and indirect transparency factors, are identified in Table \ref{table:Comprehension,Variables}. Relationships between the variable and the direct or indirect transparency factors that are not shown in Figure \ref{fig: Concept Map}, were identified after conducting correlation analyses.

\begin{table}[h]
\centering
\caption{Visualization promotion of human operator comprehension variables, relationship to the hypotheses, and the associated direct and indirect transparency factors, are presented in Figure \ref{fig: Concept Map}.}
\label{table:Comprehension,Variables}
\begin{tabular}{?l|c?c|c|c?c|c|c|c|c?}
\Cline{1pt}{3-10}
\multicolumn{1}{c}{} & & \multicolumn{8}{c?}{\textbf{Transparency Factors}} \\ \Cline{1pt}{3-10}
\multicolumn{1}{c}{} & & \multicolumn{3}{c?}{\textbf{Direct}} & \multicolumn{5}{c?}{\textbf{Indirect}} \\ \Cline{1pt}{3-10}
\multicolumn{1}{c}{} & & {\multirow[b]{6}{*}{\rotatebox{90}{\textbf{Explainability}}}} & {\multirow[b]{6}{*}{\rotatebox{90}{\textbf{Performance}}}} & {\multirow[b]{6}{*}{\rotatebox{90}{\textbf{Usability}}}} & {\multirow[b]{6}{*}{\rotatebox{90}{\textbf{Capability}}}} & {\multirow[b]{6}{*}{\rotatebox{90}{\textbf{Effectiveness}}}} & {\multirow[b]{6}{*}{\rotatebox{90}{\textbf{Justification}}}} & {\multirow[b]{6}{*}{\rotatebox{90}{\textbf{SA}}}} &  {\multirow[b]{6}{*}{\rotatebox{90}{\textbf{Understanding}}}} \\
\multicolumn{1}{c}{} & & & & & & & & & \\
\multicolumn{1}{c}{} & & & & & & & & & \\
\multicolumn{1}{c}{} & & & & & & & & & \\
\multicolumn{1}{c}{} & & & & & & & & & \\ \Cline{1pt}{1-2}
\multicolumn{1}{?c|}{\textbf{Objective Variables}} & {\textbf{Hypotheses}} & & & & & & & & \\ \Cline{1pt}{1-10}
{SA Probe Accuracy} & $H_{4}$ & & {\checkmark} & & & {\checkmark} & & {\checkmark} & \checkmark \\ \hline
{Collective Left-Clicks by SA Level} & $H_{5}$ & {\checkmark} & & {\checkmark} & & {\checkmark} & {\checkmark} & & \\ \hline
{Target Right-Clicks by SA Level} & $H_{5}$ & {\checkmark} & & {\checkmark} & & {\checkmark} & {\checkmark} & & \\ \hline
{Highest Value Target Abandoned} & $H_{4}$, $H_{5}$ & {\checkmark} & & {\checkmark} & & {\checkmark} & {\checkmark} & & \checkmark \\ \hline
{Abandoned Target Information Pop-Up Window Open} & $H_{5}$ & {\checkmark} & & {\checkmark} & & {\checkmark} & {\checkmark} & & \\ \Cline{1pt}{1-10}
\multicolumn{1}{?c}{\textbf{Subjective Variables}} & \multicolumn{9}{c?}{\textbf{}} \\ \Cline{1pt}{1-10}
{3-D Situational Awareness Rating Technique} & $H_{4}$ & & & & \checkmark &  {\checkmark} & & {\checkmark} & \checkmark \\ \Cline{1pt}{1-10}
\end{tabular}
\end{table}

Thirteen human factors display design principles, associated with perceptual operations, mental models, as well as human attention and memory \cite{Wickens2004}, suggest that information must be legible, clear, concise, organized, easily accessible, and consistent. Providing information, such as the collective state, on the collective icon, rather than using all of the individual collective entities is more clear, concise, organized, and consistent; therefore, it was hypothesized ($H_{4}$) that operators will have a better understanding of the information provided by the Collective visualization. Providing information redundantly via icons, colors, messages, and the collective and target information pop-up windows can aid operator comprehension and justify their future actions. It was hypothesized ($H_{5}$) that the Collective visualization provided information used to accurately justify actions. An ideal visualization will enable operators to perceive and comprehend information that is explainable, which will support taking accurate future actions.  

\subsection{Metrics and Results}
\label{sec: R2 metrics}

The operator had access to supplementary information that was not continually displayed, such as different colored target borders that identified which targets were in range and had been abandoned, or information pop-up windows that provided collective state and target support information, in order to aid comprehension ($SA_{2}$) of collective behavior and inform particular actions. The results of \textit{SA probe accuracy}, which is the percentage of correctly answered SA probes questions used to assess the operator's SA during a trial, identified that operators using the Collective visualization had higher SA probe accuracy, regardless of the SA level. Further details regarding the statistical tests were provided in the Metrics and Results Section \ref{section:R1 metrics} of $R_{1}$. 

\textit{Collective left-clicks} identified targets that were in range of a collective (i.e., white borders indicated that the individual collective entities were investigating the target, while yellow indicated no investigation), whether the targets had been abandoned (i.e., red borders), and was the first click required to issue a command. The number of collective left-clicks descriptive statistics 15 seconds before asking, while asking, and during response to a SA probe question are shown in Table \ref{table:Comprehension,Collective Left-Clicks}. Operators using the IA visualization had fewer collective left-clicks, regardless of when the metric was assessed for all SA levels, except for $SA_{O}$ during response to a SA probe question. The Mann-Whitney-Wilcoxon tests found highly significant effects between visualizations for $SA_{O}$ (n = 664, DOF = 1) 15 seconds before asking a SA probe question (U = 64213, $\rho$ $<$ 0.001), while asking a SA probe question (U = 67670, $\rho$ $<$ 0.001), and during response to a SA probe question (U = 64710, $\rho$ $<$ 0.001). A highly significant effect was also found when responding to a SA probe question for $SA_{2}$ (n = 223, DOF = 1, U = 8317, $\rho$ $<$ 0.001). Moderate significant effects were found for $SA_{1}$ (n = 290, DOF = 1) 15 seconds before asking a SA probe question (U = 12534, $\rho$ $<$ 0.01), while asking a SA probe question (U = 12043, $\rho$ $<$ 0.01), and during response to a SA probe question (U = 12414, $\rho$ $<$ 0.01). An additional moderate significant effect was found while asking a SA probe question for $SA_{3}$ (n = 151, DOF = 1, U = 3472, $\rho$ $<$ 0.01). Significant effects were found 15 seconds before asking a SA probe question for $SA_{2}$ (n = 223, DOF = 1, U = 7210.5, $\rho$ = 0.04) and during response to a SA probe question for $SA_{3}$ (n = 151, DOF = 1, U = 3489, $\rho$ = 0.01). No correlations were found between the number of collective left-clicks and SA probe accuracy.

\begin{table}[h]
\centering
\caption{Collective left-clicks descriptive statistics 15 seconds before asking, while asking, and during response to SA probe question by SA level.}
\label{table:Comprehension,Collective Left-Clicks}
\begin{tabular}{c|c|c|c|c|}
\cline{2-5}
 & \textbf{Timing} & \textbf{SA Level} & \textbf{Mean (SD)} & \textbf{Median (Min/Max)} \\ \hline
\multicolumn{1}{|c|}{\multirow{12}{*}{IA}} & \multirow{4}{*}{Before} & $SA_{O}$ & 1.64 (1.84) & 1 (0/12) \\ \cline{3-5} 
\multicolumn{1}{|c|}{} & & \cellcolor{light-gray}$SA_{1}$ & \cellcolor{light-gray}1.53 (1.75) & \cellcolor{light-gray}1 (0/11) \\ \cline{3-5} 
\multicolumn{1}{|c|}{} & & \cellcolor{medium-gray}$SA_{2}$ & \cellcolor{medium-gray}1.78 (1.9) & \cellcolor{medium-gray}1 (0/12) \\ \cline{3-5}
\multicolumn{1}{|c|}{} & & \cellcolor{dark-gray}$SA_{3}$ & \cellcolor{dark-gray}1.65 (1.92) & \cellcolor{dark-gray}1 (0/9) \\ \cline{2-5} 
\multicolumn{1}{|c|}{} & \multirow{4}{*}{Asking} & $SA_{O}$ & 0.49 (0.76) & 0 (0/5) \\ \cline{3-5} 
\multicolumn{1}{|c|}{} & & \cellcolor{light-gray}$SA_{1}$ & \cellcolor{light-gray}0.3 (0.6) & \cellcolor{light-gray}0 (0/3) \\ \cline{3-5}
\multicolumn{1}{|c|}{} & & \cellcolor{medium-gray}$SA_{2}$ & \cellcolor{medium-gray}0.42 (0.77) & \cellcolor{medium-gray}0 (0/4) \\ \cline{3-5}
\multicolumn{1}{|c|}{} & & \cellcolor{dark-gray}$SA_{3}$ & \cellcolor{dark-gray}0.33 (0.61) & \cellcolor{dark-gray}0 (0/3) \\ \cline{2-5} 
\multicolumn{1}{|c|}{} & \multirow{4}{*}{Responding} & $SA_{O}$ & 1.68 (1.79) & 1 (0/11) \\ \cline{3-5} 
\multicolumn{1}{|c|}{} & & \cellcolor{light-gray}$SA_{1}$ & \cellcolor{light-gray}1.14 (1.46) & \cellcolor{light-gray}1 (0/7) \\ \cline{3-5} 
\multicolumn{1}{|c|}{} & & \cellcolor{medium-gray}$SA_{2}$ & \cellcolor{medium-gray}1.46 (1.8) & \cellcolor{medium-gray}1 (0/10) \\ \cline{3-5}
\multicolumn{1}{|c|}{} & & \cellcolor{dark-gray}$SA_{3}$ & \cellcolor{dark-gray}1.53 (1.98) & \cellcolor{dark-gray}1 (1/9) \\ \hline
\multicolumn{1}{|c|}{\multirow{12}{*}{Collective}} & \multirow{4}{*}{Before} & $SA_{O}$ & 1.95 (1.57) & 2 (0/9) \\ \cline{3-5} 
\multicolumn{1}{|c|}{} & & \cellcolor{light-gray}$SA_{1}$ & \cellcolor{light-gray}1.88 (1.47) & \cellcolor{light-gray}2 (0/8) \\ \cline{3-5} 
\multicolumn{1}{|c|}{} & & \cellcolor{medium-gray}$SA_{2}$ & \cellcolor{medium-gray}2.13 (1.68) & \cellcolor{medium-gray}2 (0/9) \\ \cline{3-5}
\multicolumn{1}{|c|}{} & & \cellcolor{dark-gray}$SA_{3}$ & \cellcolor{dark-gray}1.83 (1.61) & \cellcolor{dark-gray}1 (0/7) \\ \cline{2-5} 
\multicolumn{1}{|c|}{} & \multirow{4}{*}{Asking} & $SA_{O}$ & 0.69 (0.88) & 0 (0/5) \\ \cline{3-5} 
\multicolumn{1}{|c|}{} & & \cellcolor{light-gray}$SA_{1}$ & \cellcolor{light-gray}0.51 (0.79) & \cellcolor{light-gray}0 (0/5) \\ \cline{3-5} 
\multicolumn{1}{|c|}{} & & \cellcolor{medium-gray}$SA_{2}$ & \cellcolor{medium-gray}0.91 (0.89) & \cellcolor{medium-gray}1 (0/4) \\ \cline{3-5}
\multicolumn{1}{|c|}{} & & \cellcolor{dark-gray}$SA_{3}$ & \cellcolor{dark-gray}0.73 (0.96) & \cellcolor{dark-gray}0 (0/4) \\ \cline{2-5} 
\multicolumn{1}{|c|}{} & \multirow{4}{*}{Responding} & $SA_{O}$ & 1.52 (1.21) & 1 (0/6) \\ \cline{3-5} 
\multicolumn{1}{|c|}{} & & \cellcolor{light-gray}$SA_{1}$ & \cellcolor{light-gray}1.32 (1.02) & \cellcolor{light-gray}1 (0/4) \\ \cline{3-5} 
\multicolumn{1}{|c|}{} & & \cellcolor{medium-gray}$SA_{2}$ & \cellcolor{medium-gray}1.57 (1.21) & \cellcolor{medium-gray}1 (0/5) \\ \cline{3-5}
\multicolumn{1}{|c|}{} & & \cellcolor{dark-gray}$SA_{3}$ & \cellcolor{dark-gray}1.89 (1.48) & \cellcolor{dark-gray}2 (0/6) \\ \hline
\end{tabular}
\end{table}

\textit{Target right-clicks} allowed the operator to open or close target information pop-up windows, which provided the percentage of support each collective had for a respective target. Operators may have used the support information to justify issuing commands. The number of target right-clicks descriptive statistics 15 seconds before asking, while asking, and during response to a SA probe question are presented in Table \ref{table:Comprehension,Target Right-Clicks}. The Collective visualization had fewer target right-clicks for all SA levels, 15 seconds before asking and during response to a SA probe question, while the IA visualization had fewer while asking a SA probe question. The Mann-Whitney-Wilcoxon test found no significant effects between visualizations for the number of target right-clicks. The Spearman correlation analysis revealed weak correlations between the number of target right-clicks and SA probe accuracy for the IA visualization 15 seconds before asking a SA probe question for $SA_{O}$ (r = 0.17, $\rho$ $<$ 0.01) and for $SA_{2}$ (r = 0.37, $\rho$ $<$ 0.001).

\begin{table}[h]
\centering
\caption{Target right-clicks descriptive statistics 15 seconds before asking, while asking, and during response to SA probe question by SA level.}
\label{table:Comprehension,Target Right-Clicks}
\begin{tabular}{c|c|c|c|c|}
\cline{2-5}
 & \textbf{Timing} & \textbf{SA Level} & \textbf{Mean (SD)} & \textbf{Median (Min/Max)} \\ \hline
\multicolumn{1}{|c|}{\multirow{12}{*}{IA}} & \multirow{4}{*}{Before} & $SA_{O}$ & 1.68 (2.38) & 1 (0/13) \\ \cline{3-5} 
\multicolumn{1}{|c|}{} & & \cellcolor{light-gray}$SA_{1}$ & \cellcolor{light-gray}1.92 (2.62) & \cellcolor{light-gray}1 (0/13) \\ \cline{3-5} 
\multicolumn{1}{|c|}{} & & \cellcolor{medium-gray}$SA_{2}$ & \cellcolor{medium-gray}1.28 (1.91) & \cellcolor{medium-gray}1 (0/11) \\ \cline{3-5}
\multicolumn{1}{|c|}{} & & \cellcolor{dark-gray}$SA_{3}$ & \cellcolor{dark-gray}1.8 (2.5) & \cellcolor{dark-gray}1 (0/12) \\ \cline{2-5} 
\multicolumn{1}{|c|}{} & \multirow{4}{*}{Asking} & $SA_{O}$ & 0.37 (0.79) & 0 (0/7) \\ \cline{3-5} 
\multicolumn{1}{|c|}{} & & \cellcolor{light-gray}$SA_{1}$ & \cellcolor{light-gray}0.44 (0.74) & \cellcolor{light-gray}0 (0/4) \\ \cline{3-5}
\multicolumn{1}{|c|}{} & & \cellcolor{medium-gray}$SA_{2}$ & \cellcolor{medium-gray}0.31 (0.67) & \cellcolor{medium-gray}0 (0/3) \\ \cline{3-5} 
\multicolumn{1}{|c|}{} & & \cellcolor{dark-gray}$SA_{3}$ & \cellcolor{dark-gray}0.37 (0.75) & \cellcolor{dark-gray}0 (0/3) \\ \cline{2-5} 
\multicolumn{1}{|c|}{} & \multirow{4}{*}{Responding} & $SA_{O}$ & 1.07 (1.77) & 0 (0/10) \\ \cline{3-5} 
\multicolumn{1}{|c|}{} & & \cellcolor{light-gray}$SA_{1}$ & \cellcolor{light-gray}1.11 (1.69) & \cellcolor{light-gray}0 (0/10) \\ \cline{3-5} 
\multicolumn{1}{|c|}{} & & \cellcolor{medium-gray}$SA_{2}$ & \cellcolor{medium-gray}1.1 (1.75) & \cellcolor{medium-gray}0 (0/9) \\ \cline{3-5} 
\multicolumn{1}{|c|}{} & & \cellcolor{dark-gray}$SA_{3}$ & \cellcolor{dark-gray}1.68 (2.24) & \cellcolor{dark-gray}1 (0/10) \\ \hline
\multicolumn{1}{|c|}{\multirow{12}{*}{Collective}} & \multirow{4}{*}{Before} & $SA_{O}$ & 1.52 (2.41) & 1 (0/18) \\ \cline{3-5} 
\multicolumn{1}{|c|}{} & & \cellcolor{light-gray}$SA_{1}$ & \cellcolor{light-gray}1.79 (2.71) & \cellcolor{light-gray}0 (0/18) \\ \cline{3-5} 
\multicolumn{1}{|c|}{} & & \cellcolor{medium-gray}$SA_{2}$ & \cellcolor{medium-gray}1.17 (1.94) & \cellcolor{medium-gray}0 (0/10) \\ \cline{3-5} 
\multicolumn{1}{|c|}{} & & \cellcolor{dark-gray}$SA_{3}$ & \cellcolor{dark-gray}1.49 (2.35) & \cellcolor{dark-gray}0 (0/11) \\ \cline{2-5} 
\multicolumn{1}{|c|}{} & \multirow{4}{*}{Asking} & $SA_{O}$ & 0.5 (1) & 0 (0/9) \\ \cline{3-5} 
\multicolumn{1}{|c|}{} & & \cellcolor{light-gray}$SA_{1}$ & \cellcolor{light-gray}0.49 (0.86) & \cellcolor{light-gray}0 (0/4) \\ \cline{3-5} 
\multicolumn{1}{|c|}{} & & \cellcolor{medium-gray}$SA_{2}$ & \cellcolor{medium-gray}0.55 (1.31) & \cellcolor{medium-gray}0 (0/9) \\ \cline{3-5} 
\multicolumn{1}{|c|}{} & & \cellcolor{dark-gray}$SA_{3}$ & \cellcolor{dark-gray}0.44 (0.69) & \cellcolor{dark-gray}0 (0/2) \\ \cline{2-5} 
\multicolumn{1}{|c|}{} & \multirow{4}{*}{Responding} & $SA_{O}$ & 0.99 (1.7) & 0 (0/11) \\ \cline{3-5} 
\multicolumn{1}{|c|}{} & & \cellcolor{light-gray}$SA_{1}$ & \cellcolor{light-gray}1.01 (1.74) & \cellcolor{light-gray}0 (0/11) \\ \cline{3-5} 
\multicolumn{1}{|c|}{} & & \cellcolor{medium-gray}$SA_{2}$ & \cellcolor{medium-gray}0.84 (1.44) & \cellcolor{medium-gray}0 (0/9) \\ \cline{3-5}
\multicolumn{1}{|c|}{} & & \cellcolor{dark-gray}$SA_{3}$ & \cellcolor{dark-gray}1.21 (1.98) & \cellcolor{dark-gray}0 (0/8) \\ \hline
\end{tabular}
\end{table}

The abandon command was provided to operators who desired a collective to discontinue investigating a particular target. Ideally lower valued targets were abandoned, since the objective was to aid each collective in selecting and moving to the highest valued target two sequential times. The percentage of times the \textit{highest value target was abandoned} per participant is shown in Table \ref{table:Comprehension,Highest Value Target Abandoned}. Operators using the IA visualization abandoned the highest value target less frequently, but no significant effects were found between the visualizations. 

\begin{table}[h]
\centering
\caption{The highest value target abandoned descriptive statistics per participant by decision difficulty (\%).}
\label{table:Comprehension,Highest Value Target Abandoned}
\begin{tabular}{c|c|c|c|}
\cline{2-4}
 & \textbf{Decision Difficulty} & \textbf{Mean (SD)} & \textbf{Median (Min/Max)} \\ \hline
\multicolumn{1}{|c|}{\multirow{3}{*}{IA}} & Overall & 32.36 (29.53) & 26 (0/100) \\ \cline{2-4} 
\multicolumn{1}{|c|}{} & \cellcolor{light-gray}Easy & \cellcolor{light-gray}31.2 (27.17) & \cellcolor{light-gray}25 (0/75) \\ \cline{2-4} 
\multicolumn{1}{|c|}{} & \cellcolor{medium-gray}Hard & \cellcolor{medium-gray}42.1 (40.53) & \cellcolor{medium-gray}23 (0/100) \\ \hline
\multicolumn{1}{|c|}{\multirow{3}{*}{Collective}} & Overall & 43.6 (31.94) & 38 (0/100) \\ \cline{2-4} 
\multicolumn{1}{|c|}{} & \cellcolor{light-gray}Easy & \cellcolor{light-gray}33.25 (35.96) & \cellcolor{light-gray}27 (0/100) \\ \cline{2-4} 
\multicolumn{1}{|c|}{} & \cellcolor{medium-gray}Hard & \cellcolor{medium-gray}48.75 (36.85) & \cellcolor{medium-gray}38 (0/100) \\ \hline
\end{tabular}
\end{table}

The percentage of times an \textit{abandoned target information pop-up window was open} per participant was evaluated and is represented in Table \ref{table:Comprehension,Info Window Open}. The operator may have used the support information in order to justify abandoning a particular target. Participants using the IA visualization had fewer abandoned target information pop-up windows open, compared to the Collective visualization operators. No significant effects were found between visualizations.

\begin{table}[h]
\centering
\caption{The abandoned target information pop-up window open descriptive statistics per participant by decision difficulty (\%).}
\label{table:Comprehension,Info Window Open}
\begin{tabular}{c|c|c|c|}
\cline{2-4}
 & \textbf{Decision Difficulty} & \textbf{Mean (SD)} & \textbf{Median (Min/Max)} \\ \hline
\multicolumn{1}{|c|}{\multirow{3}{*}{IA}} & Overall & 23.86 (31.43) & 10.5 (0/100) \\ \cline{2-4} 
\multicolumn{1}{|c|}{} & \cellcolor{light-gray}Easy & \cellcolor{light-gray}22.2 (30.95) & \cellcolor{light-gray}13 (0/100) \\ \cline{2-4} 
\multicolumn{1}{|c|}{} & \cellcolor{medium-gray}Hard & \cellcolor{medium-gray}28.7 (37.89) & \cellcolor{medium-gray}8.5 (0/100) \\ \hline
\multicolumn{1}{|c|}{\multirow{3}{*}{Collective}} & Overall & 33.8 (34.9) & 15 (0/100) \\ \cline{2-4} 
\multicolumn{1}{|c|}{} & \cellcolor{light-gray}Easy & \cellcolor{light-gray}30.7 (37.85) & \cellcolor{light-gray}15 (0/100) \\ \cline{2-4} 
\multicolumn{1}{|c|}{} & \cellcolor{medium-gray}Hard & \cellcolor{medium-gray}36.08 (40.87) & \cellcolor{medium-gray}14 (0/100) \\ \hline
\end{tabular}
\end{table}

The \textit{SART} results, which measured the operator's perceived situational understanding, demand on attentional resources, and supply of attentional resources \cite{Selcon1991}, were ranked higher for the Collective visualization compared to the IA. The statistical test details were provided in Section \ref{section:R1 metrics}.

\subsection{Discussion}

The analysis of how visualization promoted operator comprehension identified advantages and disadvantages associated with both visualizations. The Collective visualization promoted higher comprehension ($SA_{2}$) and situational awareness; however, because the Collective operators abandoned the highest value target more frequently, $H_{4}$ was not supported. Mistaking the roman numeral identifiers with the integer identifiers may have caused IA operator confusion and contributed to lower $SA_{2}$. Ensuring that identifiers are unique and distinct, such as integers versus letters, may mitigate misunderstanding. The target value for the Collective visualization may not have been salient enough to distinguish it from other potential targets. Further investigations are required to determine if the target value must use the entire collective hub icon area, similar to the IA visualization, in order to be recognizable, and to establish what levels of obscurity are needed to ensure that target values are distinguishable from one another. 

The use of target borders (collective left-clicks), information pop-up windows (target right-clicks), and target value, were assessed to determine if operators used these types of information to accurately justify actions. Collective right-clicks, which opened and closed collective information pop-up windows, 15 seconds before asking, while asking, and during response to a SA probe question, were not analyzed in this manuscript, because the collective evaluation did not indicate which collective was selected via the right-click. None of the metrics supported $H_{5}$, because the information provided by the Collective visualization did not justify accurate actions. Collective left-clicks did not support or hinder SA probe accuracy for either visualization, but fewer target right-clicks 15 seconds before asking a SA probe question supported higher $SA_{O}$ and $SA_{2}$ probe accuracy for operators using the IA visualization. The operators may have learned to anticipate when SA probe questions were going to be asked and took preventative actions, by opening or closing target information windows, which resulted in higher SA probe accuracy. The use of target information pop-up windows aided Collective visualization users to abandon targets more than 30\% of the time. Further analysis using technology, such as an eye-tracker, may provide more accurate metrics to determine operator comprehension during SA probe questions by identifying exactly where an operator is focusing their attention. 

\section{$R_{3}$: Visualization Usability}

Understanding \textit{which visualization promoted better usability}, $R_{3}$, is necessary to determine which system characteristics promote effective transparency in human-collective systems. The associated objective dependent variables were (1) interface clutter, (2) Euclidean distance, (3) collective and target left- and right-clicks, (4) metrics associated with abandoned targets, and (5) the time between the committed state and an issued decide command. The relationship between the variables and the corresponding hypotheses, as well as the direct and indirect transparency factors are identified in Table \ref{table:Usability,Variables}. Additional relationships that are not identified in Figure \ref{fig: Concept Map}, between the variable and the direct or indirect transparency factors are provided after conducting correlation analyses.

\begin{table}[h]
\centering
\caption{Visualization usability variables, relationship to the hypotheses, and the associated direct and indirect transparency factors, are presented in Figure \ref{fig: Concept Map}.}
\label{table:Usability,Variables}
\begin{tabular}{?l|c?c|c|c?c|c|c|c|c|c|c?} \Cline{1pt}{3-12}
\multicolumn{1}{c}{} & & \multicolumn{10}{c?}{\textbf{Transparency Factors}} \\ \Cline{1pt}{3-12}
\multicolumn{1}{c}{} & & \multicolumn{3}{c?}{\textbf{Direct}} & \multicolumn{7}{c?}{\textbf{Indirect}} \\ \Cline{1pt}{3-12}
\multicolumn{1}{c}{} & & {\multirow[b]{6}{*}{\rotatebox{90}{\textbf{Explainability}}}} & {\multirow[b]{6}{*}{\rotatebox{90}{\textbf{Performance}}}} & {\multirow[b]{6}{*}{\rotatebox{90}{\textbf{Usability}}}} & {\multirow[b]{6}{*}{\rotatebox{90}{\textbf{Effectiveness}}}} & {\multirow[b]{6}{*}{\rotatebox{90}{\textbf{Information}}}} &  {\multirow[b]{6}{*}{\rotatebox{90}{\textbf{Justification}}}} & {\multirow[b]{6}{*}{\rotatebox{90}{\textbf{Predictability}}}} & {\multirow[b]{6}{*}{\rotatebox{90}{\textbf{SA}}}} & {\multirow[b]{6}{*}{\rotatebox{90}{\textbf{Timing}}}} & {\multirow[b]{6}{*}{\rotatebox{90}{\textbf{Understanding}}}} \\
\multicolumn{1}{c}{} & & & & & & & & & & & \\ 
\multicolumn{1}{c}{} & & & & & & & & & & & \\
\multicolumn{1}{c}{} & & & & & & & & & & & \\
\multicolumn{1}{c}{} & & & & & & & & & & & \\ \Cline{1pt}{1-2}
\multicolumn{1}{?c|}{\textbf{Objective Variables}} & {\textbf{Hypotheses}} & & & & & & & & & & \\ \Cline{1pt}{1-12}
{Local Clutter Percentage} & $H_{6}$ & & {\checkmark} & {\checkmark} & {\checkmark} & & & & \checkmark & & {\checkmark} \\ \hline
{Global Clutter Percentage} & $H_{6}$ & & {\checkmark} & {\checkmark} & {\checkmark} & & & & \checkmark & & {\checkmark} \\ \hline
{Euclidean Distance Between SA Probe} & {\multirow{2}{*}{$H_{7}$}} & & & {\multirow{2}{*}{\checkmark}} & {\multirow{2}{*}{\checkmark}} & & & & & & \\ 
{Interest and Clicks} & & & & & & & & & & & \\ \hline
{Sum of Euclidean Distance Between Clicks} & $H_{7}$ & & & {\checkmark} & \checkmark & & & & & & \\ \hline
{Collective Left-Clicks per Participant} & $H_{7}$ & & & {\checkmark} & {\checkmark} & & {\checkmark} & & & & \\ \hline
{Collective Right-Clicks per Participant} & $H_{7}$ & & & {\checkmark} & {\checkmark} & {\checkmark} & {\checkmark} & & & & \\ \hline
{Target Left-Clicks per Participant} & $H_{7}$ & & & {\checkmark} & {\checkmark} & & & & & & \\ \hline
{Target Right-Clicks per Participant} & $H_{7}$ & & & {\checkmark} & {\checkmark} & \checkmark & {\checkmark} & & & & \\ \hline
{Highest Value Target Abandoned} & $H_{6}$ & {\checkmark} & & {\checkmark} & {\checkmark} & & {\checkmark} & & & & {\checkmark} \\ \hline
{Abandoned Target Information Window} & {\multirow{2}{*}{$H_{6}$}} & {\multirow{2}{*}{\checkmark}} & & {\multirow{2}{*}{\checkmark}} & {\multirow{2}{*}{\checkmark}} & & {\multirow{2}{*}{\checkmark}} & & & & \\ 
{Open} & & & & & & & & & & & \\ \hline
{Abandon Requests Exceeded Abandoned} & {\multirow{2}{*}{$H_{6}$}} & {\multirow{2}{*}{\checkmark}} & & {\multirow{2}{*}{\checkmark}} & {\multirow{2}{*}{\checkmark}} & & & & & & {\multirow{2}{*}{\checkmark}} \\ 
{Targets} & & & & & & & & & & & \\ \hline
{Time Between Committed State and Issued} & {\multirow{2}{*}{$H_{6}$}} & & & {\multirow{2}{*}{\checkmark}} & {\multirow{2}{*}{\checkmark}} & & & {\multirow{2}{*}{\checkmark}} & {\multirow{2}{*}{\checkmark}} & {\multirow{2}{*}{\checkmark}} & \\ 
{Decide Command} & & & & & & & & & & & \\ \Cline{1pt}{1-12}
\end{tabular}
\end{table}

\begin{figure}[bp!]
\begin{center}
\includegraphics[width=11.25cm]{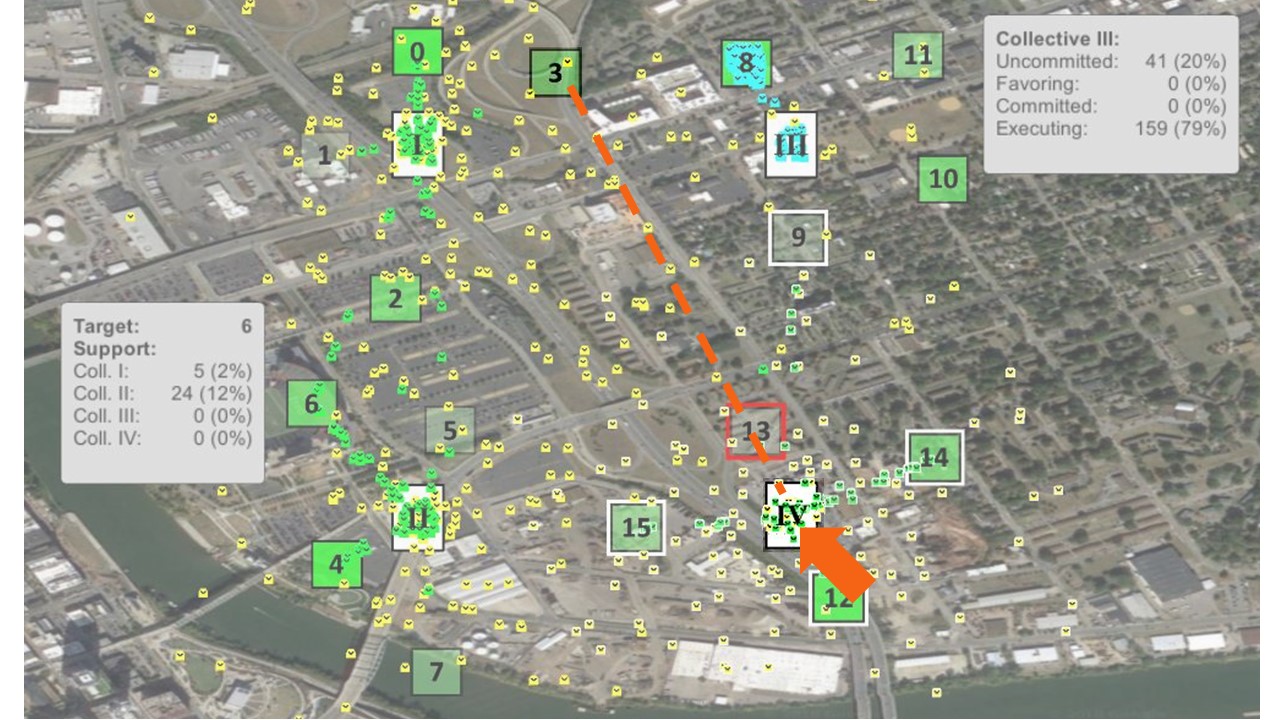}
\caption{Example of Euclidean distance between SA probe interest (Target 3) and clicks (Collective IV), denoted by an orange dashed line.}
\label{fig:Distance1}
\end{center}
\end{figure}

The goal of usability is to design systems that are effective, efficient, safe to use, have good utility, easy to learn, and are memorable \cite{Preece2007}. Ensuring good usability is necessary to ensure operators will be able to perceive and understand the information presented on a visualization, and to promote effective interactions. It was hypothesized ($H_{6}$) that the Collective visualization will promote better usability by being more predictable and explainable. Providing information that is explainable may aid operator comprehension, while predictable information may expedite operator actions. An ideal system will not require constant operator interaction to perform well; therefore, it was hypothesized ($H_{7}$) that operators using the Collective interface will require fewer interactions.

\subsection{Metrics and Results}

Many system characteristics were available to the operators in order to aid task completion. The IA visualization had both lower \textit{local clutter percentage}, which was the percentage of area obstructed by items displayed within the 500 $m$ circular radius of a collective, or target, and \textit{global clutter percentage}, which was the percentage of area obstructed by all objects displayed on the visualization. Operators using the IA visualization had fewer collective and target information pop-up windows open throughout the trial. The statistical test details were provided in Section \ref{section:R1 metrics}.

\begin{table}[bp!]
\centering
\caption{Euclidean distance between SA probe interest and clicks descriptive statistics 15 seconds before asking, while asking, and during response to SA probe question by SA level.}
\label{table:Usability,Dist 1}
\begin{tabular}{c|c|c|c|c|}
\cline{2-5}
 & \textbf{Timing} & \textbf{SA Level} & \textbf{Mean (SD)} & \textbf{Median (Min/Max)} \\ \hline
\multicolumn{1}{|c|}{\multirow{12}{*}{IA}} & \multirow{4}{*}{Before} & $SA_{O}$ & 767.1 (262.5) & 768.3 (183.4/1425.9) \\ \cline{3-5} 
\multicolumn{1}{|c|}{} & & \cellcolor{light-gray}$SA_{1}$ & \cellcolor{light-gray}759.5 (251.64) & \cellcolor{light-gray}765.1 (269/1425.9) \\ \cline{3-5} 
\multicolumn{1}{|c|}{} & & \cellcolor{medium-gray}$SA_{2}$ & \cellcolor{medium-gray}768.9 (282.07) & \cellcolor{medium-gray}787.9 (184.4/1312.4) \\ \cline{3-5}
\multicolumn{1}{|c|}{} & & \cellcolor{dark-gray}$SA_{3}$ & \cellcolor{dark-gray}783.4 (262.89) & \cellcolor{dark-gray}757.4 (183.4/1260.5) \\ \cline{2-5} 
\multicolumn{1}{|c|}{} & \multirow{4}{*}{Asking} & $SA_{O}$ & 758.44 (291.48) & 744.91 (73.72/1636.46) \\ \cline{3-5} 
\multicolumn{1}{|c|}{} & & \cellcolor{light-gray}$SA_{1}$ & \cellcolor{light-gray}754.4 (284.65) & \cellcolor{light-gray}744.6 (139.7/1636.5) \\ \cline{3-5}
\multicolumn{1}{|c|}{} & & \cellcolor{medium-gray}$SA_{2}$ & \cellcolor{medium-gray}768.4 (316.09) & \cellcolor{medium-gray}782.4 (160.2/1382.8) \\ \cline{3-5} 
\multicolumn{1}{|c|}{} & & \cellcolor{dark-gray}$SA_{3}$ & \cellcolor{dark-gray}753.7 (275.04) & \cellcolor{dark-gray}730.68 (73.72/1329.14) \\ \cline{2-5} 
\multicolumn{1}{|c|}{} & \multirow{4}{*}{Responding} & $SA_{O}$ & 764.24 (298.84) & 757.53 (73.72/1636.46) \\ \cline{3-5} 
\multicolumn{1}{|c|}{} & & \cellcolor{light-gray}$SA_{1}$ & \cellcolor{light-gray}760.9 (297.14) & \cellcolor{light-gray}746.6 (160.6/1636.5) \\ \cline{3-5} 
\multicolumn{1}{|c|}{} & & \cellcolor{medium-gray}$SA_{2}$ & \cellcolor{medium-gray}774.6 (319.08) & \cellcolor{medium-gray}777.6 (160.2/1381.2) \\ \cline{3-5} 
\multicolumn{1}{|c|}{} & & \cellcolor{dark-gray}$SA_{3}$ & \cellcolor{dark-gray}757.71 (278.14) & \cellcolor{dark-gray}749.85 (73.72/1283.9) \\ \hline
\multicolumn{1}{|c|}{\multirow{12}{*}{Collective}} & \multirow{4}{*}{Before} & $SA_{O}$ & 820.7 (255.67) & 855.9 (222.3/1470.5) \\ \cline{3-5} 
\multicolumn{1}{|c|}{} & & \cellcolor{light-gray}$SA_{1}$ & \cellcolor{light-gray}825.6 (264.1) & \cellcolor{light-gray}828.1 (249.8/1470.5) \\ \cline{3-5} 
\multicolumn{1}{|c|}{} & & \cellcolor{medium-gray}$SA_{2}$ & \cellcolor{medium-gray}812.9 (234.94) & \cellcolor{medium-gray}865.2 (317.4/1329.9) \\ \cline{3-5} 
\multicolumn{1}{|c|}{} & & \cellcolor{dark-gray}$SA_{3}$ & \cellcolor{dark-gray}821.6 (271.03) & \cellcolor{dark-gray}873.1 (222.3/1243.9) \\ \cline{2-5} 
\multicolumn{1}{|c|}{} & \multirow{4}{*}{Asking} & $SA_{O}$ & 851.4 (293.91) & 859.5 (280.2/1745.2) \\ \cline{3-5} 
\multicolumn{1}{|c|}{} & & \cellcolor{light-gray}$SA_{1}$ & \cellcolor{light-gray}845.5 (282.53) & \cellcolor{light-gray}862.5 (281.5/1745.2) \\ \cline{3-5} 
\multicolumn{1}{|c|}{} & & \cellcolor{medium-gray}$SA_{2}$ & \cellcolor{medium-gray}879.5 (299.93) & \cellcolor{medium-gray}869.2 (280.2/1696.1) \\ \cline{3-5} 
\multicolumn{1}{|c|}{} & & \cellcolor{dark-gray}$SA_{3}$ & \cellcolor{dark-gray}823.5 (314.47) & \cellcolor{dark-gray}787.6 (314.7/1469) \\ \cline{2-5} 
\multicolumn{1}{|c|}{} & \multirow{4}{*}{Responding} & $SA_{O}$ & 827.7 (273.83) & 819.8 (258.1/1546) \\ \cline{3-5} 
\multicolumn{1}{|c|}{} & & \cellcolor{light-gray}$SA_{1}$ & \cellcolor{light-gray}827.9 (279.21) & \cellcolor{light-gray}819 (366.6/1484.6) \\ \cline{3-5} 
\multicolumn{1}{|c|}{} & & \cellcolor{medium-gray}$SA_{2}$ & \cellcolor{medium-gray}845.2 (275.55) & \cellcolor{medium-gray}862.7 (258.1/1546) \\ \cline{3-5}
\multicolumn{1}{|c|}{} & & \cellcolor{dark-gray}$SA_{3}$ & \cellcolor{dark-gray}799.7 (261.1) & \cellcolor{dark-gray}797.3 (314.7/1378.8) \\ \hline
\end{tabular}
\end{table}

The Euclidean \textit{distance (pixels) between the focus of the SA probe question and where the operator was interacting} with the visualization indicated where operators focused their attention, because no eye-tracker was used. Euclidean distance can be used to assess the effectiveness of the object placements on the display. Larger distances are not ideal, because more time \cite{Gillan1992} and effort is required to locate and interact with the object. The first requirement of calculating the Euclidean distance was to determine what the collective, or target of interest was in a SA probe question. For example, Target 3 is the target of interest for the following question: ``What collectives are investigating Target 3?'' The second requirement was to determine where the operator was interacting with the system. The Euclidean distance between Target 3 and the operator's current interaction (i.e., click), which was Collective IV, is identified by a dashed orange line in Figure \ref{fig:Distance1}. The Euclidean distance between SA probe interest and clicks descriptive statistics 15 seconds before asking, while asking, and during response to a SA probe question are presented in Table \ref{table:Usability,Dist 1}. Operators using the IA visualization had shorter Euclidean distances between the SA probe interest and their clicks with the visualization, regardless of when the metric was assessed for all SA levels. The Mann-Whitney-Wilcoxon tests found a moderate significant effect between visualizations while asking a SA probe question for $SA_{O}$ (n = 464, DOF = 1, U = 31052, $\rho$ $<$ 0.01). Highly significant effects were found between visualizations 15 seconds before asking a SA probe question for $SA_{O}$ (n = 557, DOF = 1, U = 43303, $\rho$ = 0.02) and for $SA_{1}$ (n = 273, DOF = 1, U = 10577, $\rho$ = 0.05), while asking a SA probe question for $SA_{1}$ (n = 229, DOF = 1, U = 7645, $\rho$ = 0.01), and during response to a SA probe for $SA_{O}$ (n = 499, DOF = 1, U = 35029, $\rho$ = 0.02). The Spearman correlation analysis revealed a weak correlation between the Euclidean distance between the SA probe's focus interest and the clicks and SA probe accuracy for the IA visualization 15 seconds before asking a SA probe question for $SA_{1}$ (r = -0.18, $\rho$ = 0.04).

The \textit{sum Euclidean distance (pixels) between clicks} was the sum of all distances between the operator's current interaction and the immediately previous interaction. For example, if an operator interacted with the visualization four times while a SA probe question was being asked, the sum Euclidean distance is the sum between interactions one and two, interactions two and three, and interactions three and four. The sum of Euclidean distance between clicks descriptive statistics 15 seconds before asking, while asking, and during response to a SA probe question are presented in Table \ref{table:Usability,Dist 2}. Operators using the Collective visualization had smaller sums of Euclidean distance between their interactions, regardless of when the metric was assessed for all SA levels, with two exceptions. The IA visualization had a smaller sum for $SA_{1}$ while asking and during response to a SA probe question. The Mann-Whitney-Wilcoxon test found no significant effects between visualizations. The Spearman correlation analysis revealed weak correlations between the sum of Euclidean distance between clicks and SA probe accuracy for the IA visualization 15 seconds before asking a SA probe question for $SA_{2}$ (r = 0.2, $\rho$ = 0.04) and during response to a SA probe question for $SA_{O}$ (r = 0.14, $\rho$ = 0.02). Weak correlations were also revealed for the Collective visualization while asking a SA probe question for $SA_{O}$ (r = -0.13, $\rho$ = 0.05) and for $SA_{1}$ (r = -0.2, $\rho$ = 0.05). 

\begin{table}[h]
\centering
\caption{Sum of Euclidean distance between clicks descriptive statistics 15 seconds before asking, while asking, and during response to SA probe question by SA level.}
\label{table:Usability,Dist 2}
\begin{tabular}{c|c|c|c|c|}
\cline{2-5}
 & \textbf{Timing} & \textbf{SA Level} & \textbf{Mean (SD)} & \textbf{Median (Min/Max)} \\ \hline
\multicolumn{1}{|c|}{\multirow{12}{*}{IA}} & \multirow{4}{*}{Before} & $SA_{O}$ & 3386 (2911.78) & 2500 (0/21115) \\ \cline{3-5} 
\multicolumn{1}{|c|}{} & & \cellcolor{light-gray}$SA_{1}$ & \cellcolor{light-gray}3185 (2665.35) & \cellcolor{light-gray}2476 (0/18465) \\ \cline{3-5} 
\multicolumn{1}{|c|}{} & & \cellcolor{medium-gray}$SA_{2}$ & \cellcolor{medium-gray}3389 (3091.22) & \cellcolor{medium-gray}2469 (0/21115) \\ \cline{3-5}
\multicolumn{1}{|c|}{} & & \cellcolor{dark-gray}$SA_{3}$ & \cellcolor{dark-gray}3732.6 (3071.18) & \cellcolor{dark-gray}2901.7 (246.1/17323.6) \\ \cline{2-5} 
\multicolumn{1}{|c|}{} & \multirow{4}{*}{Asking} & $SA_{O}$ & 1748.9 (1779.6) & 1250.7 (0/12161.7) \\ \cline{3-5} 
\multicolumn{1}{|c|}{} & & \cellcolor{light-gray}$SA_{1}$ & \cellcolor{light-gray}1387.3 (1328.49) & \cellcolor{light-gray}985.7 (0/6777.4) \\ \cline{3-5}
\multicolumn{1}{|c|}{} & & \cellcolor{medium-gray}$SA_{2}$ & \cellcolor{medium-gray}2030.1 (2053.98) & \cellcolor{medium-gray}1477.5 (0/12161.7) \\ \cline{3-5} 
\multicolumn{1}{|c|}{} & & \cellcolor{dark-gray}$SA_{3}$ & \cellcolor{dark-gray}1975 (1954.76) & \cellcolor{dark-gray}1362 (0/8949) \\ \cline{2-5} 
\multicolumn{1}{|c|}{} & \multirow{4}{*}{Responding} & $SA_{O}$ & 1383.2 (1416.84) & 934.8 (0/9315.1) \\ \cline{3-5} 
\multicolumn{1}{|c|}{} & & \cellcolor{light-gray}$SA_{1}$ & \cellcolor{light-gray}1088.7 (1108.12) & \cellcolor{light-gray}793.5 (0/4999) \\ \cline{3-5} 
\multicolumn{1}{|c|}{} & & \cellcolor{medium-gray}$SA_{2}$ & \cellcolor{medium-gray}1606 (1610.13) & \cellcolor{medium-gray}1310 (0/9315) \\ \cline{3-5} 
\multicolumn{1}{|c|}{} & & \cellcolor{dark-gray}$SA_{3}$ & \cellcolor{dark-gray}1582.6 (1534.29) & \cellcolor{dark-gray}1211.9 (0/6728.2) \\ \hline
\multicolumn{1}{|c|}{\multirow{12}{*}{Collective}} & \multirow{4}{*}{Before} & $SA_{O}$ & 3187 (1983.72) & 2834 (0/9690) \\ \cline{3-5} 
\multicolumn{1}{|c|}{} & & \cellcolor{light-gray}$SA_{1}$ & \cellcolor{light-gray}3085 (1911.3) & \cellcolor{light-gray}2611 (0/8898) \\ \cline{3-5} 
\multicolumn{1}{|c|}{} & & \cellcolor{medium-gray}$SA_{2}$ & \cellcolor{medium-gray}3235 (2049.38) & \cellcolor{medium-gray}3041 (0/9690) \\ \cline{3-5} 
\multicolumn{1}{|c|}{} & & \cellcolor{dark-gray}$SA_{3}$ & \cellcolor{dark-gray}3331.6 (2047.61) & \cellcolor{dark-gray}3097.7 (195.8/8910.7) \\ \cline{2-5} 
\multicolumn{1}{|c|}{} & \multirow{4}{*}{Asking} & $SA_{O}$ & 1741.3 (1323.11) & 1434.6 (0/6323.2) \\ \cline{3-5} 
\multicolumn{1}{|c|}{} & & \cellcolor{light-gray}$SA_{1}$ & \cellcolor{light-gray}1778 (1437) & \cellcolor{light-gray}1380 (0/5920) \\ \cline{3-5} 
\multicolumn{1}{|c|}{} & & \cellcolor{medium-gray}$SA_{2}$ & \cellcolor{medium-gray}1693 (1093.21) & \cellcolor{medium-gray}1619 (0/4411) \\ \cline{3-5} 
\multicolumn{1}{|c|}{} & & \cellcolor{dark-gray}$SA_{3}$ & \cellcolor{dark-gray}1743.8 (1439.54) & \cellcolor{dark-gray}1371 (0/6323.2) \\ \cline{2-5} 
\multicolumn{1}{|c|}{} & \multirow{4}{*}{Responding} & $SA_{O}$ & 1218.5 (944.41) & 1047.1 (0/4428.2) \\ \cline{3-5} 
\multicolumn{1}{|c|}{} & & \cellcolor{light-gray}$SA_{1}$ & \cellcolor{light-gray}1174.7 (992.56) & \cellcolor{light-gray}995 (0/4277.6) \\ \cline{3-5} 
\multicolumn{1}{|c|}{} & & \cellcolor{medium-gray}$SA_{2}$ & \cellcolor{medium-gray}1176.8 (801.24) & \cellcolor{medium-gray}1074.6 (0/4084.9) \\ \cline{3-5}
\multicolumn{1}{|c|}{} & & \cellcolor{dark-gray}$SA_{3}$ & \cellcolor{dark-gray}1373.15 (1041) & \cellcolor{dark-gray}1160.81 (65.35/4428.24) \\ \hline
\end{tabular}
\end{table}

Collective and target left- and right-clicks were examined per participant. \textit{Target left-clicks} were the second click required in the process of issuing commands, but did not provide supplementary information. The number of collective and target left- and right-clicks descriptive statistics are presented in Table \ref{table:Usability,Left Right-Clicks}. Operators using the IA visualization had fewer collective and target left-clicks, while those using the Collective visualization had fewer collective and target right-clicks. The Mann-Whitney-Wilcoxon test identified no significant effects between visualizations.

\begin{table}[h]
\centering
\caption{Number of collective and target left- and right-clicks per participant descriptive statistics.}
\label{table:Usability,Left Right-Clicks}
\begin{tabular}{c|c|c|c|}
\cline{2-4}
 & \textbf{Clicks} & \textbf{Mean (SD)} & \textbf{Median (Min/Max)} \\ \hline
\multicolumn{1}{|c|}{\multirow{4}{*}{IA}} & Collective Left & 107.6 (49.89) & 104 (5/235) \\ \cline{2-4} 
\multicolumn{1}{|c|}{} & \cellcolor{light-gray}Collective Right & \cellcolor{light-gray}30.64 (20.98) & \cellcolor{light-gray}27.5 (0/85) \\ \cline{2-4} 
\multicolumn{1}{|c|}{} & \cellcolor{medium-gray}Target Left & \cellcolor{medium-gray}97.64 (58.78) & \cellcolor{medium-gray}83 (5/251) \\ \cline{2-4} 
\multicolumn{1}{|c|}{} & \cellcolor{dark-gray}Target Right & \cellcolor{dark-gray}97.18 (82.79) & \cellcolor{dark-gray}68.5 (4/352) \\ \hline
\multicolumn{1}{|c|}{\multirow{4}{*}{Collective}} & Collective Left & 121.96 (47.4) & 130.5 (35/212) \\ \cline{2-4} 
\multicolumn{1}{|c|}{} & \cellcolor{light-gray}Collective Right & \cellcolor{light-gray}30.57 (31.95) & \cellcolor{light-gray}19.5 (7/164) \\ \cline{2-4} 
\multicolumn{1}{|c|}{} & \cellcolor{medium-gray}Target Left & \cellcolor{medium-gray}185.6 (64.32) & \cellcolor{medium-gray}202 (62/290) \\ \cline{2-4}
\multicolumn{1}{|c|}{} & \cellcolor{dark-gray}Target Right & \cellcolor{dark-gray}82.39 (60.22) & \cellcolor{dark-gray}75 (23/278) \\ \hline
\end{tabular}
\end{table}

Metrics showing how operators used the abandon command were assessed. IA operators had lower percentages of times the \textit{highest value target was abandoned} and lower percentages of times an \textit{abandoned target information pop-up window was open} per participant. The statistical analyses of both metrics were provided in Section \ref{sec: R2 metrics}. Instances may have occurred when the operator accidentally issued an undesired abandon command or repeatedly issued the abandon command, although the command only needed to be issued once; hence, the percent of times \textit{abandon commands exceeded abandoned targets} was examined and the descriptive statistics are shown in Table \ref{table:Usability,Abandon Exceeded}. Operators using the IA visualization had fewer repeated abandon commands for all decision difficulties, compared to those using the Collective visualization. The Mann-Whitney-Wilcoxon test found no significant effects between visualizations.

\begin{table}[h]
\centering
\caption{The percentage of times abandon commands exceeded abandoned targets per participant descriptive statistics.}
\label{table:Usability,Abandon Exceeded}
\begin{tabular}{c|c|c|c|}
\cline{2-4}
 & \textbf{Decision Difficulty} & \textbf{Mean (SD)} & \textbf{Median (Min/Max)} \\ \hline
\multicolumn{1}{|c|}{\multirow{3}{*}{IA}} & Overall & 1.18 (3.02) & 0 (0/12) \\ \cline{2-4} 
\multicolumn{1}{|c|}{} & \cellcolor{light-gray}Easy & \cellcolor{light-gray}0.4 (1.55) & \cellcolor{light-gray}0 (0/6) \\ \cline{2-4} 
\multicolumn{1}{|c|}{} & \cellcolor{medium-gray}Hard & \cellcolor{medium-gray}1.35 (4) & \cellcolor{medium-gray}0 (0/16) \\ \hline
\multicolumn{1}{|c|}{\multirow{3}{*}{Collective}} & Overall & 2.68 (6.27) & 0 (0/22) \\ \cline{2-4} 
\multicolumn{1}{|c|}{} & \cellcolor{light-gray}Easy & \cellcolor{light-gray}2.05 (5.06) & \cellcolor{light-gray}0 (0/15) \\ \cline{2-4} 
\multicolumn{1}{|c|}{} & \cellcolor{medium-gray}Hard & \cellcolor{medium-gray}3.08 (7.74) & \cellcolor{medium-gray}0 (0/25) \\ \hline
\end{tabular}
\end{table}

The \textit{time difference (minutes) between the committed state and issued decide command} assessed the operator's ability to predict the collective's future state changing from the committed state (30\% support for a target) to executing (50\% support for a target). The time difference between the committed state and when an operator issued a decide command descriptive statistics are shown in Table \ref{table:Usability,Timing of Commit}. Operators using the Collective visualization had smaller time differences between the committed state and issued decide commands for overall and easy decisions; however, operators using the IA visualization had smaller time differences for hard decisions. The Mann-Whitney-Wilcoxon test found no significant effects between visualizations. 

\begin{table}[h]
\centering
\caption{The time difference (minutes) between committed state and issued decide request per participant descriptive statistics.}
\label{table:Usability,Timing of Commit}
\begin{tabular}{c|c|c|c|}
\cline{2-4}
 & \textbf{Decision Difficulty} & \textbf{Mean (SD)} & \textbf{Median (Min/Max)} \\ \hline
\multicolumn{1}{|c|}{\multirow{3}{*}{IA}} & Overall & 0.68 (0.27) & 0.62 (0.42/1.78) \\ \cline{2-4} 
\multicolumn{1}{|c|}{} & \cellcolor{light-gray}Easy & \cellcolor{light-gray}0.7 (0.47) & \cellcolor{light-gray}0.63 (0.32/2.56) \\ \cline{2-4} 
\multicolumn{1}{|c|}{} & \cellcolor{medium-gray}Hard & \cellcolor{medium-gray}0.72 (0.21) & \cellcolor{medium-gray}0.66 (0.41/1.15) \\ \hline
\multicolumn{1}{|c|}{\multirow{3}{*}{Collective}} & Overall & 0.65 (0.15) & 0.63 (0.45/1.18) \\ \cline{2-4} 
\multicolumn{1}{|c|}{} & \cellcolor{light-gray}Easy & \cellcolor{light-gray}0.56 (0.14) & \cellcolor{light-gray}0.58 (0.27/0.88) \\ \cline{2-4} 
\multicolumn{1}{|c|}{} & \cellcolor{medium-gray}Hard & \cellcolor{medium-gray}0.78 (0.3) & \cellcolor{medium-gray}0.75 (0.47/1.99) \\ \hline
\end{tabular}
\end{table}

\subsection{Discussion}

The analysis of which visualization promoted better usability was inconclusive, because both visualizations had advantages and disadvantages. The Collective visualization operators were able to predict and issue decide commands faster, after the collective was in the committed state, compared to those using the IA visualization. $H_{6}$ was not supported; however, because the Collective visualization operators abandoned the highest value target more frequently, and there was a higher percentage of abandon commands exceeding abandoned targets. Operators using the Collective visualization had more local and global clutter, which suggests that Collective operators relied on the information pop-up windows to answer the SA probe questions more than IA operators. Sixteen of twenty-four SA probe questions relied on information provided in the information-pop up windows. The collective and target icons, as well as the target outlines, were intended to aid Collective operators to answer the SA probe questions correctly; however, the operators needed to use the information pop-up windows in order to see the numeric values for the collective support and collective behavior and answer questions regarding target support from a specific collective, or multiple collectives. An example question, such as ``What collectives are investigating Target 3?" in Figure \ref{fig: Collective Interface}, will require using the target information pop-up window, because Target 3 is in range of Collective I and III. A target information pop-up window is not required for Target 1, in Figure \ref{fig: Collective Interface}, since it is only in range of Collective I. The need to use information pop-up windows contributed to the Collective visualization clutter. The operators may have preferred the numeric value representations versus the other visualization techniques, which may have contributed to their reliance on the information pop-up windows. The IA operators may have had an advantage, by deducing the same information as the Collective operators gained from the information pop-up windows, by observing the dynamic behavior of the individual collective entities. Relying on supplemental information pop-up windows is not ideal and suggests that improvements must be made to the collective icon to ensure the collective's state information is more understandable. Modifications, such as indicating which collective was the highest supporting collective on the target icon, instead of solely showing that there was support through the use of color and opacity, may increase the reliance on the target icon instead of the target information pop-up window. Additional experimental design modifications can ensure a more even distribution of questions that may rely on other information providing features on the visualization, such as the icons, system messages, or collective assignments versus information pop-up windows. Operators using target information pop-up windows to verifying that a target was abandoned by a collective, may have been confused if the reported target support was greater than zero. The operators may have reissued additional abandon commands in an attempt to reduce the collective support to zero, although only one abandon command was needed. There were instances during the trial when a few individual collective entities became lost, when the collective hub was transitioning to a new location, and never moved with the hub. The lost entities may have continued to explore a now abandoned target, because they never received the message to abandon the target, which occurred inside of the hub. Strategies, such as reporting zero percent support when an abandon command is issued, may help mitigate erroneous repeated abandon command behavior, which was experienced up to 25\% for some operators. Operators using the IA visualization may have also experienced confusion if they saw individual collective entities still travelling to an abandoned target. Not displaying lost entities after a specific period of time once a collective hub has moved to a new location may also reduce the number of reissued abandon commands. 

$SA_{1}$ probe questions that inquired about objects nearby were answered more accurately than those that were further away when using the IA visualization. Asking SA probe questions about objects at various distances from the operator's current focal point is necessary in order to understand how clutter, or moving individual collective entities, may affect the operator's ability to identify the object of interest and answer the question correctly. Smaller sum of Euclidean distances between interactions, suggests Collective operators may have had fewer interactions. Further analysis is required to determine whether more interactions were needed for operators to answer SA probe questions correctly and make more successful and faster decisions. $H_{7}$ was not supported by the analysis. The IA operators may have issued fewer commands, or did not rely on target borders as much as operators using the Collective visualization. Issuing more commands suggests that Collective operators may have wanted more control over the decision-making task, which may have occurred due to lower trust, or misunderstanding collective behavior. Further investigations are needed in order to understand how the algorithm and control mechanisms may interact with the visualization of collectives to influence operator behavior. The effectiveness of the system design will be dependent on all system characteristics working together to promote optimal human-collective performance and to achieve high effectiveness. 

\section{$R_{4}$: Visualization Influence on Human-Collective Performance}

Assessing \textit{which visualization promoted better human-collective performance}, $R_{4}$, is necessary to determine whether the human-collective system transparency aided task completion. An ideal system performs a task quickly, safely, and successfully. The associated objective dependent variables were (1) decision time, (2) selection success rate, and (3) SA probe accuracy. The relationship between the variables and the corresponding hypotheses, as well as the direct and indirect transparency factors, are identified in Table \ref{table:Performance,Variables}. Additional relationships between the variable and the direct or indirect transparency factors, not identified in Figure \ref{fig: Concept Map}, are provided due to correlation analyses.

\begin{table}[h]
\centering
\caption{Visualization influence on human-collective performance variables, relationship to the hypotheses, and the associated direct and indirect transparency factors, are presented in Figure \ref{fig: Concept Map}.}
\label{table:Performance,Variables}
\begin{tabular}{?l|c?c?c|c|c|c|c|c?}
\Cline{1pt}{3-9}
\multicolumn{1}{c}{} & & \multicolumn{7}{c?}{\textbf{Transparency Factors}} \\ \Cline{1pt}{3-9}
\multicolumn{1}{c}{} & & {\textbf{Direct}} & \multicolumn{6}{c?}{\textbf{Indirect}} \\ \Cline{1pt}{3-9}
\multicolumn{1}{c}{} & & {\multirow[b]{6}{*}{\rotatebox{90}{\textbf{Performance}}}} & {\multirow[b]{6}{*}{\rotatebox{90}{\textbf{Capability}}}} & {\multirow[b]{6}{*}{\rotatebox{90}{\textbf{Effectiveness}}}} & {\multirow[b]{6}{*}{\rotatebox{90}{\textbf{Efficiency}}}} & {\multirow[b]{6}{*}{\rotatebox{90}{\textbf{SA}}}} &  {\multirow[b]{6}{*}{\rotatebox{90}{\textbf{Timing}}}} &  {\multirow[b]{6}{*}{\rotatebox{90}{\textbf{Understanding}}}} \\
\multicolumn{1}{c}{} & & & & & & & & \\ 
\multicolumn{1}{c}{} & & & & & & & & \\
\multicolumn{1}{c}{} & & & & & & & & \\
\multicolumn{1}{c}{} & & & & & & & & \\ \Cline{1pt}{1-2}
\multicolumn{1}{?c|}{\textbf{Objective Variables}} & {\textbf{Hypotheses}} & & & & & & & \\ \Cline{1pt}{1-9}
{Decision Time Per Decision} & $H_{8}$ & {\checkmark} & & {\checkmark} & {\checkmark} & & {\checkmark} & \\ \hline
{Selection Success Rate Per Decision} & $H_{8}$ & {\checkmark} & & {\checkmark} & & & & \\ \hline
{SA Probe Accuracy} & $H_{8}$ & {\checkmark} & & {\checkmark} & & {\checkmark} & & \checkmark \\  \hline
{Mental Rotation Assessment} & $H_{8}$ & {\checkmark} & {\checkmark} & & & & & \\  \Cline{1pt}{1-9}
\multicolumn{1}{?c}{\textbf{Subjective Variables}} & \multicolumn{8}{c?}{\textbf{}} \\ \Cline{1pt}{1-9}
{Weekly Hours on a Desktop or Laptop} & $H_{8}$ & {\checkmark} & {\checkmark} & & & & & \\ \hline
{Video Game Proficiency} & $H_{8}$ & {\checkmark} & {\checkmark} & & & & & \\ \hline
{Post-Trial Performance and Understanding} & $H_{8}$ & {\checkmark} & \checkmark & & & & & \checkmark \\ \Cline{1pt}{1-9}
\end{tabular}
\end{table}

Performance of the human-collective team can be used to assess the effects of visualization transparency on the team's ability to fulfill tasks. An ideal system design desires high performance rates. It was hypothesized ($H_{8}$) that the human-collective performance, effectiveness, efficiency, and timing will be better using the Collective visualization. 

\subsection{Metrics and Results}

\begin{table}[bp!]
\centering
\caption{Decision time (minutes) descriptive statistics per decision difficulty.}
\label{table:Performance,Decision Time}
\begin{tabular}{c|c|c|c|c|}
\cline{2-5}
 & \textbf{Model} & \textbf{Decision Difficulty} & \textbf{Mean (SD)} & \textbf{Median (Min/Max)} \\ \hline
\multicolumn{1}{|c|}{\multirow{6}{*}{IA}} & \multirow{3}{*}{$M_{2}$} & Overall & 4.32 (1.83) & 3.94 (1.74/15.94) \\ \cline{3-5} 
\multicolumn{1}{|c|}{} & & \cellcolor{light-gray}Easy & \cellcolor{light-gray}3.77 (1.63) & \cellcolor{light-gray}3.38 (1.74/13.47) \\ \cline{3-5} 
\multicolumn{1}{|c|}{} & & \cellcolor{medium-gray}Hard & \cellcolor{medium-gray}5.09 (1.82) & \cellcolor{medium-gray}4.68 (1.86/15.94) \\ \cline{2-5} 
\multicolumn{1}{|c|}{} & \multirow{3}{*}{$M_{2 SIM}$} & Overall & 4.8 (1.1) & 4.82 (2.46/7.68) \\ \cline{3-5} 
\multicolumn{1}{|c|}{} & & \cellcolor{light-gray}Easy & \cellcolor{light-gray}4.19 (1.06) & \cellcolor{light-gray}4.07 (2.46/8.85) \\ \cline{3-5} \multicolumn{1}{|c|}{} & & \cellcolor{medium-gray}Hard & \cellcolor{medium-gray}5.73 (1.26) & \cellcolor{medium-gray}5.54 (3.43/10.15) \\ \hline
\multicolumn{1}{|c|}{\multirow{6}{*}{Collective}} & \multirow{3}{*}{$M_{2}$} & Overall & 3.97 (1.37) & 3.64 (1.83/9.94) \\ \cline{3-5} 
\multicolumn{1}{|c|}{} & & \cellcolor{light-gray}Easy & \cellcolor{light-gray}3.37 (1.23) & \cellcolor{light-gray}3.09 (1.83/9.94) \\ \cline{3-5} 
\multicolumn{1}{|c|}{} & & \cellcolor{medium-gray}Hard & \cellcolor{medium-gray}4.67 (1.2) & \cellcolor{medium-gray}4.57 (2.46/8.81) \\ \cline{2-5} 
\multicolumn{1}{|c|}{} & \multirow{3}{*}{$M_{2 SIM}$} & Overall & 4.79 (1.11) & 4.79 (2.49/7.7) \\ \cline{3-5} 
\multicolumn{1}{|c|}{} & & \cellcolor{light-gray}Easy & \cellcolor{light-gray}4.17 (0.93) & \cellcolor{light-gray}4.1 (2.49/7.55) \\ \cline{3-5}
\multicolumn{1}{|c|}{} & & \cellcolor{medium-gray}Hard & \cellcolor{medium-gray}5.77 (1.38) & \cellcolor{medium-gray}5.62 (3.67/10.25) \\ \hline
\end{tabular}
\end{table}

The length of time it took the human-collective team to reach a decision, \textit{decision time} (minutes), was examined. Consensus decision-making algorithms are inherently slow, which is undesirable in realistic use scenarios. Adding a human operator into the loop permits the human to influence the decision and has the potential to minimize decision time. The decision time descriptive statistics per decision are shown in Table \ref{table:Performance,Decision Time} \cite{Roundtree20191}. Operators using the Collective visualization had faster decision times for overall, easy, and hard decisions. Both visualizations had faster human-collective decision times compared to the simulation. The Collective visualization simulation had slightly faster decision times for overall and easy decisions, while the IA visualization simulation had faster decision times for hard decisions. The Mann-Whitney-Wilcoxon test found significant effects between visualizations with human operators for overall (n = 672, DOF = 1, U = 50921, $\rho$ = 0.03), easy (n = 375, DOF = 1, U = 15452, $\rho$ = 0.04), and hard decisions (n = 297, DOF = 1, U = 9521, $\rho$ = 0.04). Highly significant effects were found between human operators and simulation for the IA visualization overall (n = 672, DOF = 1, U = 74005, $\rho$ $<$ 0.001), easy (n = 481, DOF = 1, U = 37414, $\rho$ $<$ 0.001), and hard decisions (n = 384, DOF = 1, U = 23194, $\rho$ $<$ 0.001) and for the Collective visualization overall (n = 672, DOF = 1, U = 79468, $\rho$ $<$ 0.001), easy (n = 461, DOF = 1, U = 38786, $\rho$ $<$ 0.001), and hard decisions (n = 392, DOF = 1, U = 26887, $\rho$ $<$ 0.001).

The \textit{selection success rate} was the number of correct decisions (the collective moved to the highest valued target) relative to the total number of decisions. Selection success rate descriptive statistics per decision are shown in Table \ref{table:Performance,Success Rate} \cite{Roundtree20191}. Collective operators had higher selection success rates for all decision difficulties. The human-collective teams had higher selection success rates for both visualizations for all decision difficulties compared to the simulation. The Collective simulation had higher selection success rates for overall and hard decisions, while the IA simulation had higher selection success rates for easy decisions. The Mann-Whitney-Wilcoxon test found highly significant effects between visualizations with human operators for overall (n = 672, DOF = 1, U = 64008, $\rho$ $<$ 0.001) and easy decisions (n = 375, DOF = 1, U = 19845, $\rho$ $<$ 0.001). A moderate significant effect significant effects between visualizations with human operators was found for hard decisions (n = 297, DOF = 1, U = 12761, $\rho$ $<$ 0.01). Highly significant effects were found between human operators and simulation for the IA visualization overall (n = 672, DOF = 1, U = 34650, $\rho$ $<$ 0.001), easy (n = 481, DOF = 1, U = 18242, $\rho$ $<$ 0.001), and hard decisions (n = 384, DOF = 1, U = 13088, $\rho$ $<$ 0.001) and for the Collective visualization overall (n = 672, DOF = 1, U = 22162, $\rho$ $<$ 0.001), easy (n = 461, DOF = 1, U = 10795, $\rho$ $<$ 0.001), and hard decisions (n = 392, DOF = 1, U = 9449.5, $\rho$ $<$ 0.001).

\begin{table}[h]
\centering
\caption{Selection success rate (\%) descriptive statistics per decision difficulty.}
\label{table:Performance,Success Rate}
\begin{tabular}{c|c|c|c|c|}
\cline{2-5}
 & \textbf{Model} & \textbf{Decision Difficulty} & \textbf{Mean (SD)} & \textbf{Median (Min/Max)} \\ \hline
\multicolumn{1}{|c|}{\multirow{6}{*}{IA}} & \multirow{3}{*}{$M_{2}$} & Overall & 75 (43.37) & 100 (0/100) \\ \cline{3-5} 
\multicolumn{1}{|c|}{} & & \cellcolor{light-gray}Easy & \cellcolor{light-gray}81.44 (38.98) & \cellcolor{light-gray}100 (0/100) \\ \cline{3-5} 
\multicolumn{1}{|c|}{} & & \cellcolor{medium-gray}Hard & \cellcolor{medium-gray}66.2 (47.47) & \cellcolor{medium-gray}100 (0/100) \\ \cline{2-5} 
\multicolumn{1}{|c|}{} & \multirow{3}{*}{$M_{2 SIM}$} & Overall & 73.69 (19.01) & 70 (20/100) \\ \cline{3-5} 
\multicolumn{1}{|c|}{} & & \cellcolor{light-gray}Easy & \cellcolor{light-gray}77.15 (27.12) & \cellcolor{light-gray}85.71 (0/100) \\ \cline{3-5} \multicolumn{1}{|c|}{} & & \cellcolor{medium-gray}Hard & \cellcolor{medium-gray}62.05 (27.17) & \cellcolor{medium-gray}66.67 (0/100) \\ \hline
\multicolumn{1}{|c|}{\multirow{6}{*}{Collective}} & \multirow{3}{*}{$M_{2}$} & Overall & 88.39 (32.08) & 100 (0/100) \\ \cline{3-5} 
\multicolumn{1}{|c|}{} & & \cellcolor{light-gray}Easy & \cellcolor{light-gray}94.44 (22.97) & \cellcolor{light-gray}100 (0/100) \\ \cline{3-5} 
\multicolumn{1}{|c|}{} & & \cellcolor{medium-gray}Hard & \cellcolor{medium-gray}81.41 (39.03) & \cellcolor{medium-gray}100 (0/100) \\ \cline{2-5} 
\multicolumn{1}{|c|}{} & \multirow{3}{*}{$M_{2 SIM}$} & Overall & 74.58 (18.39) & 70 (20/100) \\ \cline{3-5} 
\multicolumn{1}{|c|}{} & & \cellcolor{light-gray}Easy & \cellcolor{light-gray}76.7 (26.87) & \cellcolor{light-gray}83.33 (0/100) \\ \cline{3-5}
\multicolumn{1}{|c|}{} & & \cellcolor{medium-gray}Hard & \cellcolor{medium-gray}64.04 (26.38) & \cellcolor{medium-gray}66.67 (0/100) \\ \hline
\end{tabular}
\end{table}

The Spearman correlation analysis revealed a moderate correlation between human-collective team decision time and selection success rate using the IA visualization for easy decisions (r = -0.42, $\rho$ $<$ 0.001). Weak correlations were revealed between the human-collective team decision time and selection success rate using the IA visualization for overall decisions (r = -0.27, $\rho$ $<$ 0.001) and when using the Collective visualization for overall (r = -0.11, $\rho$ = 0.05), easy (r = -0.18, $\rho$ = 0.02), and hard decisions (r = 0.18, $\rho$ = 0.03). Moderate correlations were revealed between simulation decision time and selection success rate using the IA visualization for overall decisions (r = -0.53, $\rho$ $<$ 0.001) and when using the Collective visualization for overall (r = -0.57, $\rho$ $<$ 0.001) and easy decisions (r = -0.44, $\rho$ $<$ 0.001). Weak correlations were revealed between simulation decision time and selection success rate using the IA visualization for easy (r = -0.35, $\rho$ $<$ 0.001) and hard decisions (r = -0.14, $\rho$ = 0.03).

\textit{SA probe accuracy}, which is the percentage of correctly answered SA probes questions used to assess the operator's SA during a trial, results identified that Collective operators had higher SA probe accuracy, regardless of the SA level. Further details about the statistical tests were provided in Section \ref{section:R1 metrics}.

Additional Spearman correlation analyses were conducted to see if any correlations were identified between the weekly hours that participants' used a desktop or laptop, video game proficiency, the mental rotations assessment, and selection success rate. A weak correlation was found between weekly hours participants' used a desktop or laptop and selection success rate for the IA visualization (r = 0.16, $\rho$ = 0.02).

The post-trial questionnaire assessed the participants' \textit{understanding of the collective behavior}, never (1) to always (7), and their \textit{ability} to chose the best target for each decision, never (1) to always (7). The post-trial performance and understanding subjective ranking descriptive statistics are presented in Table \ref{table:Performance,PT Performance} \cite{Cody2018}. Performance and understanding rankings were higher for operators using the Collective visualization. The Mann-Whitney-Wilcoxon test found a significant effect between visualizations for understanding (n = 56, DOF = 1, U = 513, $\rho$ = 0.04).

\begin{table}[h]
\centering
\caption{Post-trial performance and understanding model ranking descriptive statistics (1-low, 7-high).}
\label{table:Performance,PT Performance}
\begin{tabular}{c|c|c|c|}
\cline{2-4}
 & \textbf{Metric} & \textbf{Mean (SD)} & \textbf{Median (Min/Max)} \\ \hline
\multicolumn{1}{|c|}{\multirow{2}{*}{IA}} & Performance & 5.25 (1.69) & 6 (2/7) \\ \cline{2-4} 
\multicolumn{1}{|c|}{} & \cellcolor{light-gray}Understanding & \cellcolor{light-gray}4.89 (1.75) & \cellcolor{light-gray}5 (1/7) \\ \hline
\multicolumn{1}{|c|}{\multirow{2}{*}{Collective}} & Performance & 5.54 (1.29) & 6 (3/7) \\ \cline{2-4} 
\multicolumn{1}{|c|}{} & \cellcolor{light-gray}Understanding & \cellcolor{light-gray}5.82 (1.16) & \cellcolor{light-gray}6 (3/7) \\ \hline 
\end{tabular}
\end{table}

\subsection{Discussion}

The analysis suggests that the Collective visualization promoted better human-collective performance. $H_{8}$ was supported, because the Collective visualization produced higher objective primary and secondary task performance, as well as higher subjective performance. Operators using the Collective visualization had significantly faster decision times and higher selection success rates compared to the IA visualization. Realistic human-collective user scenarios will require high performance in short decision times, especially in proposed high-risk environments. Longer decision times contributed to higher success rates for both visualizations for overall decisions and for easy decisions for the IA visualization; however, faster decision times contributed to higher selection success for hard decisions using the Collective visualization. The design of an effective human-collective system must enable the human-collective team to fulfill primary objectives, without hindering other metrics, such as decision time. Devoting more time to ensure high task performance is a common trade-off behavior observed in teams. Expedited decisions may have occurred if higher valued targets were further away from other objects (less clutter), making them more salient, or if impatient operators influenced collective behaviors more to make decisions faster. The Collective visualization enabled operators with different capabilities to perform relatively the same, unlike the IA visualization, which found that individuals with more weekly desktop or laptop exposure had higher selection success rates. 

\section{Discussion}

The research objective was to determine which visualization achieved better transparency, identify what metrics were useful in determining better transparency, and to create design guidance for human-collective systems. The analysis indicated that the Collective visualization provided better transparency, because operators with different individual differences performed similarly for both the primary and secondary tasks, and the human-collective team performed better. The Mental Rotations Assessment, NASA-TLX, and SART were useful individual operator capability metrics when determining the influence of the visualization on the operator and can be easily used in other collective evaluations. Operator experience (e.g., weekly hours on a desktop or laptop) and expertise (e.g., video game proficiency) can indicate the desired operator knowledge in order to interact with the collective system effectively. The influence of visualization on human operator comprehension and visualization usability requires further investigation in order to better understand the influence of operator interactions and identify more reliable metrics to assess operator understanding. Using correlations between an operator's interactions and SA probe accuracy can be used to inform designers whether the actions taken by operators aided their responses, but does not necessarily provide insight regarding comprehension. The use of eye-tracking technology can provide improved insight regarding operator comprehension and usability by recording where the operator was looking 15 seconds before asking, while asking, and during response to a SA probe question. Where operators are looking on the visualization prior to taking action will indicate what types of information the operator was potentially perceiving and comprehending, the difficulty to identify the desired information due to visualization clutter, and the duration of time devoted to looking at particular information. Clutter will greatly impact an operator's ability to perceive and comprehend information on the visualization and is an informative metric to use when assessing transparency for human-collective systems.

The reliance on collective and target information pop-up windows contributed to the majority of clutter for both visualizations and suggests alterations to design recommendations for future human-collective systems. Providing supplementary information, via information pop-up windows, is necessary for successful human-collective behavior; however, other strategies must be implemented to improve the efficacy of the collective icon. Indicating which collective state is most supported, by displaying either U, F, C, or X, instead of the status of all four states, may be more advantageous to the operator. The target a collective is favoring, committed, or executing may also be displayed on the collective icon. For example, if a collective is favoring Target 8, the collective icon can show F8, which stands for Favoring Target 8. Providing the most supported state and target may enable the operator to quickly understand what the collective is doing and determine if interventions, or more support is needed to ensure successful decision making. Showing the predominant collective state; however, requires the operator to remember what U, F, C, and X stand for, which can be mitigated by adding this information to the legend. Bolding the predominant state letter on the current collective icon can be used as an alternative design change to improve the operator's understanding of the most supported collective state. The perceived mental demand and effort associated with the Collective visualization may decrease with strategies that make the collective state more obvious to the operator. 

The dynamic and streamlined behavior of the individual collective entities on the IA visualization may have aided operator understanding of collective behavior, mitigating the need to access as many information pop-up windows. Displaying all of the individual collective entities is not ideal as collectives become larger in size. The increased number of entities will contribute to more clutter, potentially hindering the operator's perception and comprehension. Using iconography, such as arrows pointing in the direction of the most supported target, or providing predicted hub locations, may be design strategies that can improve operator understanding for abstract collective visualizations. Further analysis is needed to verify the effectiveness of the proposed strategies. The consistent improvement in decision time and selection success rate across all decision difficulties suggests that using visualizations that show all the individual collective entities does not contribute to better human-collective performance. The Collective visualization enabled better human-collective performance, which is valuable as collective systems become more complex, with improved capabilities and the utilization of heterogeneous collectives. Presenting individual collective entities may have caused more stress, or confusion, and required operators to slow down the collective decision-making process in order to attain higher selection success rates. SA probe accuracy, selection success rates, and decision times were useful metrics that can be easily used in other collective system evaluations.   

Indicating target value through the use of color and opacity was a successful transparency implementation and metric to assess. Further analysis is required to determine if the entire target icon must represent the target value to be more perceivable, since operators using the Collective visualization, which used half of the target icon to represent the value and the other half to represent support from the highest supporting collective, abandoned the highest value target more frequently. Opacity levels must also be validated to ensure distinction of low-, medium-, and high-valued targets from one another. Reiterating the task objective, to choose and move each collective to the highest value target for each decision, numerous times during training may help mitigate operator misunderstanding. Improvements can be made to the collective and target information pop-up windows in order to decrease the number of reissued abandon commands for operators who relied and verified abandonment using the information pop-up windows. When an abandon command is issued the corresponding target information pop-up window can immediately report zero support, instead of the actual collective support, which corresponds closer to operators' mental models. A red bar can be overlaid on the particular collective that abandoned the target as a secondary measure to ensure the operator understands the collective status. Operators can focus their attention on collective support values that are not highlighted in red, in case the current target is the highest value target for another nearby collective. Using metrics, such as the number of times abandoned requests exceed abandoned targets, can be used as an error metric for collective systems.

Design guidance recommendations, provided in Table \ref{table:Discussion,Design}, were created from the analysis for human-collective system visualizations supporting a best-of-n sequential decision-making task. The recommendations are applicable irrespective of visualization type. The majority of the recommendations indicate that the use of color to distinguish objects from one another, or convey particular information, can be useful, as long as the operator's cognitive capacity is not exceeded. Other design strategies, such as the use of patterns, may need to be considered if operators are color blind. Several recommendations suggest providing particular information, such as collective behaviors and state information, operator actions, and system messages, in order to facilitate operator understanding. Further investigations are needed in order to determine the effectiveness of the following design recommendations for real-world scenarios where bandwidth limitations occur. Understanding how information latency and inaccurate collective state information negatively influence human-collective behavior is essential to design a resilient transparent visualization. 

\begin{table}[h]
\centering
\caption{Human-collective visualization design guidance.}
\label{table:Discussion,Design}
\begin{tabular}{|l|}
\hline
\multicolumn{1}{|c|}{\textbf{Design Guidance}} \\ \hline
1. Provide detailed supplemental information to the operator, such as the use of information pop-up windows. \\ \hline
2. Provide information about the system and operator actions, such as the use of system messages and collective \\ 
assignments windows. \\ \hline
3. The use of color and different opacity is an effective method of conveying a non-numeric value. \\ \hline
4. The use of colored borders is an effective method of distinguishing objects in the environment. \\ \hline
5. Provide a legend detailing information in order to alleviate memory demands of the operator. \\ \hline
6. Use distinct and unique identifiers for objects in the environment, such as integers versus letters. \\ \hline
7. Provide information about collective behavior that coincides with operator mental models of operation, such as \\
abandoning a target will result in zero individual collective entity support. \\ \hline
8. Indicate the status of operator commands, such as a red indicator to denote completion of a command and \\ 
green to denote an ongoing state. \\ \hline
9. Provide the projected state of a collective, such as a dynamic moving border. \\ \hline
10. Limit the number of colors used to seven plus or minus two, which is consistent with human cognitive \\ 
capacity \cite{Miller1994}. \\ \hline
11. Provide perceivable collective state information, such as the use of color to denote different states. \\ \hline
\end{tabular}
\end{table}

Visualization transparency for human-collective systems can be achieved via different design strategies and must be assessed holistically by understanding how the different factors impact transparency and are influenced by transparency. The four secondary research questions assessed four categories of transparency factors that contribute to an effective system: 1) different operator individual capabilities, 2) operator comprehension , 3) visualization usability, and 4) human-collective team performance. An ideal visualization will enable operators with different individual capabilities to perform relatively the same, promote operator comprehension, be usable, and promote high human-collective performance. The Collective visualization enabled operators with different individual capabilities to perform relatively the same and promoted better human-collective performance. The IA visualization enabled operators to perceive collective behaviors and collective support for particular targets more readily than the Collective visualization, where operators used the collective and target information pop-up windows to affirm these types of collective behaviors. As collective systems grow in size, visualizations that show all of the individual collective entities will cause perceptual and comprehension challenges, as well as influence operator actions negatively, because too many individual collective entities will clutter the display. The same advantageous observation (i.e., dynamically seeing collective behaviors and support) from this analysis may not occur with large collectives ($>$ 10000). Abstract collective visualizations designed using the provided guidelines may help promote better transparency, than visualizations showing all of the individual collective entities and enable effective human-collective teams. 

\section{Conclusion}

Designers of human-collective systems continue to debate what visualization is needed to represent collective systems better and provide transparency of the collective behaviors to the operators. This manuscript evaluates a traditional and abstract collective visualization for a sequential best-of-\textit{n} decision-making task with four collectives, each consisting of 200 individual collective entities. The visualization transparency is evaluated with respect to how the visualization impacts the human operators, operator comprehension, usability, and human-collective performance. The Collective visualization was considered more transparent, because operators with varying individual differences and capabilities were able to perform similarly in both the primary and secondary tasks, and the human-collective performance was higher compared to the IA visualization. Additional interaction analysis and metrics are needed to evaluate which visualization promoted better operator comprehension and usability. Designers must build collective systems that are effective regardless of how large the collective size may become, how simple or complex the collective behaviors are, and how real-world use scenarios, such as bandwidth limitations, may impact the system. Visualizations that show all of the individual collective entities will challenge operators' ability to perceive and comprehend information in order to inform actions. Abstract collective visualizations may be more resilient to real-world scenarios, and provide transparency to enable effective human-collective teams. 

\section*{Acknowledgements}

ONR Award N000141613025 and the US Office of Naval Research Awards N000141210987 and N00014161302 supported this effort. The work of Jason R. Cody was fully supported by the United States Military Academy and the United States Army Advanced Civil Schooling program. 








\end{document}